\documentclass[reprint,amsmath,amssymb,aps,prb,citeautoscript]{revtex4-1}
\usepackage{graphicx}
\usepackage{dcolumn}
\usepackage{bm}
\usepackage{bbold}
\usepackage{siunitx}
\usepackage{natbib}
\usepackage{verbatim}
\usepackage[T2A,T1]{fontenc}

\usepackage{soul}

\usepackage[colorlinks,linkcolor=black,anchorcolor=blue,citecolor=blue,urlcolor=blue]{hyperref}

\begin{document}
% NICK: a nice compact title, carrying the pitfalls of the word superlubricity
\title{Directional superlubricity}

% Other possible title, on the line discussed in the car from the Dead Sea to Rehovot:
\title{How matched/mismatched can two crystalline surfaces be? The geometric conditions for high friction or superlubricity}
% AS: or this one as well formulated on the same trip:
\title{Design principles for crystalline interfaces: directional locking, directional superlubricity and structural lubricity}
%idea born @ ICTP...
\title{Frictionless nanohighways}
%Just a joke
\title{Frictionless nanohighways and where to find them}
%Xin: I add one more choice
\title{Frictionless nanohighways on crystalline surfaces}

\author{Emanuele Panizon$^{1,5,\dagger}$}
\author{Andrea Silva$^{2,3,\dagger}$}
\author{Xin Cao$^{1}$}
\author{Jin Wang$^{3}$}
\author{Clemens Bechinger$^{1}$}
\author{Andrea Vanossi$^{2,3}$}
\author{Erio Tosatti$^{3,5,2}$}
\author{Nicola Manini$^{4}$}
\email{nicola.manini@unimi.it}
\affiliation{$^1$Fachbereich Physik, University Konstanz, 78464 Konstanz, Germany}
\affiliation{$^2$CNR-IOM, Consiglio Nazionale delle Ricerche - Istituto Officina dei Materiali, c/o SISSA, 34136 Trieste, Italy}
\affiliation{$^3$International School for Advanced Studies (SISSA), Via Bonomea 265, 34136 Trieste, Italy}
\affiliation{$^4$Dipartimento di Fisica, Universit\`a degli Studi di Milano, Via Celoria 16, 20133 Milano, Italy}
\affiliation{$^5$International Centre for Theoretical Physics (ICTP), Strada Costiera 11, 34151 Trieste, Italy}
\date{\today}
\thanks{Emanuele Panizon and Andrea Silva contributed equally}

% From Sci. Adv. guidelines
%"The abstract provides a snapshot of the research in the paper and should be no more than 150 words in a single paragraph written for a general readership. Do not include citations. Provide a sentence offering a general introduction to the field and then a sentence with more detailed background specific to the research described. Follow this with a very brief explanation of objectives/methods and then key results. The final sentence should describe the main conclusions of the research. Any abbreviations that appear in the title should be defined in the abstract. Graphic representations of the abstract are not permitted."
\begin{abstract}
The understanding of friction at nano-scales, ruled by the regular arrangement of atoms, is surprisingly incomplete.
Here we provide a unified understanding by studying the interlocking potential energy of two infinite contacting surfaces with arbitrary lattice symmetries, and extending it to finite contacts.
We categorize, based purely on geometrical features, all possible contacts into three different types: a structurally
lubric contact where the monolayer can move isotropically without friction, a corrugated and strongly
interlocked contact, and a newly discovered directionally structurally lubric contact where the layer can
move frictionlessly along one specific direction and retains finite friction along all other directions.
This novel category is energetically stable against rotational perturbations and provides extreme
friction anisotropy. The finite-size analysis shows that our categorization applies to a wide range
of technologically relevant materials in contact, from adsorbates on crystal surfaces to layered two-dimensional materials and colloidal monolayers.
\end{abstract}

% AS: alt version of 150 words:
% Moving objects at the nanoscale follow frictional laws unfamiliar to us, governed by the regular arrangement of surface atoms. Current knowledge of the corrugation at crystalline surfaces is still incomplete.  Here we provide a unified understanding by studying the interlocking potential energy of two infinite, arbitrary crystalline surfaces in contact and extending it to finite size. We categorize, based purely on geometrical features, all possible contacts into three different types: structurally lubric contact, where the monolayer can move frictionlessly isotropically, a strongly interlocked contact, and a newly discovered directionally structural lubric contact, where the layer can move frictionlessly along one specific direction and retains finite friction along all others. This novel category is energetically stable against rotational perturbations and provides extreme friction anisotropy. The finite-size analysis shows that our categorization applies to a wide range of materials in contact, from adsorbates on crystal surfaces to two-dimensional materials and colloidal monolayers.
\maketitle

% AS: approx word count now: 6300 (Nat. Comm. ~5000 Sci. Adv. ~10000)
\section{Introduction}

%\ep{
When two macroscopically rough surfaces are brought close to each other, they interact only locally at the touching asperities. The progressive increase of touching asperities with load at constant nominal area leads to a contact friction that is proportional to load while independent of area.
Contact friction at the atomic scale, on the other hand, follows remarkably different rules~\cite{Mo2009, Guerra2010, Kawai2016,liu2020negative,he1999adsorbed}.
At small length scales, static friction is largely dependent on the atomic arrangement, and, specifically for crystalline materials, on the mutual relation between the lattice periodicities of the two contacting surfaces.
When two atomically flat surfaces come close to each other, the atoms or molecules on one surface can fall into the interatomic gaps of another, leading to an interlocking potential that is strongly dependent on the atomic arrangement at the two contacting surfaces. Such atomic interlocking potentials have strong influences on many nanoscopic frictional processes such as scanning tunneling microscopy experiments~\cite{hirano1997, Feng2013},
%nanomanipulations~\cite{dietzel2008, Gigli2017,chen2019chemical}
nanomanipulations~\cite{dietzel2008,Gigli2018peeling,Silva2022peeling}
and fabrications of layered two dimensional (2D) materials \cite{novoselov20162d}.
However, our knowledge of such interlocking effects is incomplete and often obtained in a case-by-case manner, largely due to the material-dependent properties and the complexities of the contacting interfaces such as the contact incommensurabilities and %evidence of
strong finite-size effects.
%}

%\xc{When bringing two rough surfaces close to each other, they begin to interlock at the touching asperities which leads to a friction against relative motion.
%Contact friction at the atomic scale follows remarkably different rules compared to what we experience in our everyday life in the macroscopic world~\cite{Mo2009, Guerra2010, Kawai2016, smith2013quantized}. At nanoscopic length scales, friction is largely dependent on the atomic arrangement, and, specifically for crystalline materials, on the mutual relation between the lattice periodicities of the two contacting surfaces.
%that is proportional to the normal load but independent of contact area.
%Similarly, when two atomically flat surfaces come close to each other, the atoms or molecules on one surface can fall into the interatomic gaps of another, leading to an interlocking potential that is strongly dependent on the atomic arrangement at the two contacting surfaces. Such atomic interlocking potential have strong influences on many nanoscopic frictional processes such as scanning probe microscopy experiments, nanomanipulations, fabrications of layered two dimensional mateirals, surface based catalyst and epitaxial growth. However, our knowledge of such interlocking effects are incomplete and are often obtained in a case-by-case manner, largely due to the material-dependent properties and the complexities of the contacting interfaces such as the contact incommensurabilities and evidence of strong finite-size effects.} %Xin: I try to write a start of the introduction, feel free to change/correct or to remove.}

In an exemplary commensurate contact where the atoms on one surface can perfectly %overlap
match
the inter-atom gaps
of the other, interlocking effects are strong as the forces required to unlock individual atoms add up, thus producing a total friction that grows linearly with contact size.
On the other hand, when the contacting periodicities are incommensurate, or not perfectly aligned, interlocking forces cancel out, %possibly
tendentially
leading to
superlubric sliding or structural superlubricity~\cite{Hod2018,Vanossi2020a,Song2018RobustSuperlub, baykara2018emerging}.
The relative orientation and directions of motion of a rigid crystalline cluster interacting with an underlying periodic surface were recently shown to be dominated by the possible emergence of common lattice vectors (CLVs) in real and in reciprocal space~\cite{Cao2019a, Cao2021}.
Despite results for specific geometries and potentials~\cite{deWijn2012, Dietzel2013,hod2013registry}, a precise and general quantitative classification of the degree and type of commensurability of contacting lattices and its connection to friction is still currently called for. %incomplete.
The present paper aims at filling this gap.

In concrete, we answer the following question: given a crystal-crystal interface that involves two 2D lattices, what kind of
ideal
frictional behavior can we predict?
For a 1D interface, idealized in the Frenkel-Kontorova model~\cite{Braun1998}, the answer to this question depends on whether the lattice-spacing ratio $a/b$ is rational (high friction) or irrational (superlubric sliding for sufficiently rigid crystals).
For the 2D geometry at hand, we classify three very different friction regimes that can arise depending on the (in)commensuration detail in the relation between the two lattices.
In between the standard fully commensurate and fully incommensurate conditions, we identify and characterize an %novel ERIO: really novel? didn't we say above that it was encountered before by Cao? AS: in that case it was regarded as directional locking, here we show it scales as a superlubric contact, this was never reported before.
intermediate situation, involving an effective 1D-commensuration that produces a finite friction in one direction, but leaves incommensurate superlubric sliding in the perpendicular direction.
This condition leads to the ``locked''
%\textcolor{red}{ERIO: why locked? see below.. Perhaps we should say "geometrically determined"?}
free-sliding direction which in the title we refer to as a {\em frictionless nanohighway}, a case of
{\em directional structural lubricity}.
This phenomenon arises when two surfaces lock
%\textcolor{red}{ERIO: the word "lock" applies if that special direction, besides being superlubric, is also a sharp local energy minimum. Is that the case?
%AS: Yes, that's the beauty of it! See Fig. 5.}
into a specific orientation, such that static friction vanishes in one specific direction, while remaining finite in all other directions.

Relaxing the condition of infinite size, we obtain theoretical bounds -- plus numerical and experimental evidence -- of finite-size contacts where
approximate versions of each of the infinite-size regimes are realized.
In these quasi-commensurate contacts the friction forces along different directions grow as a function of the contact size  following very different scaling laws: sublinear and linear in low- and high-friction directions respectively.
The emerging anisotropy of friction effectively reproduces the main features of the infinite-size limit also for real-life finite-size contacts.

The results obtained in this paper are derived for rigid lattices.
The geometric conditions to differentiate the different friction regimes represents necessary -- but possibly not sufficient -- conditions for real (i.e.\ elastic) systems~\cite{benassi2015breakdown,Sharp2016}.
% NICK: well done! Of course AV Erio etc. could suggest more papers appropriate to cite here...}
%In a numerical example we will actually relax the %rigid
%assumption of rigidity, and in the discussion we comment on the limits of this assumption.
%
%ERIO: Cryptic statement, the last one! What does it mean? Why put it here?
%EMA: Is it clearer now? We could remove it, it is just if we want to introduce the fact that we do everything without elasiticity.
% AS: I update it again, referring to the somewhat-elastic simulations by Jin and the discussion in the end.
% NICK: somewhat-elastic indeed!  I'm not sure it is worth mentioning that simulation here in the introduction.

\section{The geometric context}
\begin{figure*}
 \centering
\includegraphics[width=\textwidth]{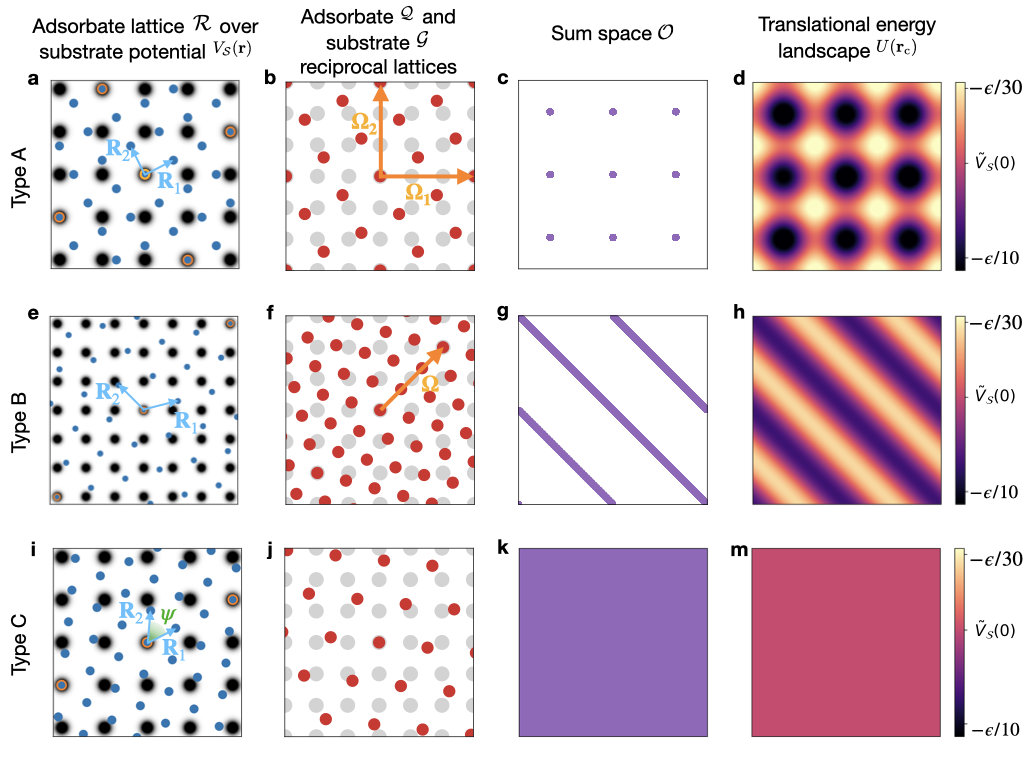}
\caption{
The three types of contacts between two 2D lattices. %\xc{Xin: The title of the figure should be bold in font, if we submit to Nat Comm or Sci Adv.}
(a--d) fully commensurate, type A; (e--h) 1D-commensurate, type B; (i--m) fully incommensurate, type C.
Panels (a, e, i) display the two contacting lattices in real space: grayscale map of the potential $V_\mathcal{S}(\mathbf{r}_\text{c})$ with unit-side %\xc{(?) AS: side of unit 1, I think it's fairly understandable}
square-lattice periodicity (black to white = low to high energy);
blue dots = adsorbate crystal;
orange-edge blue dots = CLVs in real space;
light blue arrows = the primitive vectors of the adsorbate lattice.
%%%%%
(b, f, j) the corresponding two dual lattices: red dots = adsorbate reciprocal lattice $\mathcal Q$; gray dots = substrate reciprocal lattice $\mathcal G$; arrows =
%the primitive vectors of the coincidence lattice}. % AS: ambiguous, needs not to be a lattice, it's a set.
coincidence vectors (elements of the set $\Omega$).
(c, g, k) the linear sum $\mathcal{O}$ of the
$\mathcal{R}$ and $\mathcal{S}$
lattices in the Wigner-Seitz cell of the substrate.
%\xc{as defined in } \as{Section \ref{sec:lattice_algebra}}. NICK: all stuff shown here is defined in Sec. III, not just the linear sum....
(d, h, m) the resulting interlocking potential energy (Eq.~\eqref{UofrDef}, colorbar at the right) as a function of the adsorbate translation
within the substrate Wigner-Seitz cell; for the adopted
%Gaussian % Gaussian is just an ingredient!, The potential whose average is the value given below is the full periodic potential V_S...
corrugation potential
%defined in
of
Eq.~\eqref{vperiodic}, $\tilde{V}_\mathcal{S}(\mathbf{0})=-0.0628\epsilon$.
%%%%%%%
%\xc{I would move the following part to merge with the beginning of the caption. AS: we need to define the rows and the column. Now is first column and then rows. It's matter of taste I think, but I am not sure I would re-arrange the whole caption at this point.
%NICK: No, keep it here.  I had tried to have this at the beginning, but the caption becomes too messy... }
The adsorbate lattices are:
(a--d) a square lattice with $|a| = \sqrt{5}/3$, and orientation $\theta_\mathrm{o} = \tan^{-1}(1/2)$;
(e--h) a triangular lattice with $|a| = \sqrt{3/2}$, and orientation $\theta_\mathrm{o} = 15^\circ$;
(i--m) a rhombic lattice with primitive vectors of length $|a| = \sqrt{5}/3$ separated by an angle $\psi = 1~\text{rad}$, with orientation
$\theta_\mathrm{o} = \text{tan}^{-1}(1/2)$.
}
\label{fig:coverages}
\end{figure*}

We start by recalling a few useful results in the framework of the algebra of reciprocal (dual) lattices, which will allow us to provide a full classification of the contacts between rigid periodic solids.

Consider the problem of moving a monolayer adsorbate
crystal across another crystalline surface.
%\xc{Each adsorbate atom interact with the underlying crystal surface via a periodic potential} AS: this is not the interaction between atoms and the surface, but potential generated by the surface itself: indeed it is defined at any position r, not just at the atoms ones. I think it's more clear to clearly separate the substrate potential (defined by the other cristalline interface and the interlocking potential, which indeed arises when we start to consider the how the adsorbate atoms interact with the underlying crystal. Makes sense? Also we want to put the accent on the periodicity of V_S here.
%which generates a lateral corrugation  potential $V_{\mathcal{S}}(\mathbf{r})$
Each adsorbate atom interacts with the lateral corrugation potential generated by the underlying crystal surface $V_{\mathcal{S}}(\mathbf{r})$, characterized by the following periodicity:
$V_{\mathcal{S}}(\mathbf{r})=V_{\mathcal{S}}(\mathbf{r}+n_1\mathbf{S}_1+n_2\mathbf{S}_2)$,
%induced by a second lattice with primitive vectors
where $\mathbf{S}_1$ and $\mathbf{S}_2$ are primitive vectors of the underlying crystal lattice.
For
example, we
%\xc{simplicity}, we
% AS: mmm, it's not really for simplicity, we could do this with any other function or construct it in a different way, e.g. sum of palne wave and obtain the sinusoidal function (which doesn't require the sum over the lattice points). Here we chose this to compare with previous publications.
%can
construct $V_{\mathcal{S}}$ as
\begin{equation}\label{vperiodic}
    V_\mathcal{S}(\mathbf{r})=\sum_{n_1,n_2} V(|\mathbf{r}+n_1\mathbf{S}_1+n_2\mathbf{S}_2|)
\end{equation}
where
$V(r) = -\epsilon \exp(-r^2/\sigma^2)$, with $\sigma=0.1 \, |\mathbf{S}_1|$.
Figure \ref{fig:coverages}a,e,i illustrates the crystal overlayered on a square-symmetry substrate potential.
The surface forces that contrast the motion of a rigid overlayer result from the gradient of the total (interlocking) potential energy. %, of which we consider the
Its value normalized per monolayer particle is %, namely
\begin{equation} \label{UofrDef}
U(\mathbf{r}_\mathrm{c}) = \frac 1N \sum_{j=1}^N
V_\mathcal{S}(\mathbf{r}_\mathrm{c}+\mathbf{R}_j)
\,.
\end{equation}
Here $\mathbf{r}_\mathrm{c}$ is the center-of-mass displacement and $\mathbf{R}_j$ are a set of $N$ lattice-translation vectors, generated as integer linear combinations of two primitive lattice vectors $\mathbf{R}_1$ and $\mathbf{R}_2$ of the adsorbate.
Precisely the list of the $N$ translations defines the contact shape and size.

As derived previously~\cite{Cao2019a,Cao2021}, in the $N\to \infty$ limit of a macroscopically large perfect crystalline monolayer, the interlocking potential is
\begin{equation} \label{Uofr}
U(\mathbf{r}_\mathrm{c}) = \sum_{\mathbf{\Omega}}
\tilde{V}_\mathcal{S}(\mathbf{\Omega})
\exp(i\mathbf{\Omega} \cdot  \mathbf{r}_\mathrm{c})
\,.
\end{equation}
Here, the $\mathbf{\Omega}$s are %\xc{intersections, or}
coincidence lattice vectors (CLVs) common to
%\xc{(existing in?)} % AS: they exists in R and S, regarless of Omega. I think commont to is the right mathematical term here.
both reciprocal lattices of the monolayer and the periodic surface.  $\tilde{V}_\mathcal{S}(\mathbf{\Omega})$ are the relevant Fourier components of $V_\mathcal{S}(\mathbf{r})$, defined as:
\begin{equation}
\tilde{V}_\mathcal{S}(\mathbf{G}) := \frac 1{A_\mathcal{S}} \int_{A_\mathcal{S}} V_\mathcal{S}(\mathbf{r})
\,e^{-i\mathbf{G}\cdot \mathbf{r}}\,
\text{d}^2\mathbf{r}
\,,
\end{equation}
where $A_\mathcal{S}$ is the area of the primitive cell of $\mathcal{S}$, $i$ is the imaginary unit.
Expression~\eqref{Uofr} is a 2D generalization of the well-known ``Poisson summation formula'', $\sum_{n} s(n) = \sum_{k} \tilde{s}(k)$
where $\tilde{s}$ is the Fourier transform of $s$, and the summations run from ${-\infty}$ to ${\infty}$. The dependence of the potential $U(\mathbf{r}_\mathrm{c})$ upon the relative orientation of the two contacting lattices enters the expression in Eq.~\eqref{Uofr} implicitly via the CLVs $\mathbf{\Omega}$.

Notice that the $\mathbf{\Omega}$'s included in the summation %\xc{could reach infinite size, i.e. even a large Fourier components can contribute significantly to the interlocking potential energy. This is in contrast to}
% AS: not correct. The sum always have infinite size, but with a sinusoidal potential most of these are zero.
have unlimited size.
A hypothetical purely sinusoidal potential $V_{\mathcal S}({\mathbf r})$, as commonly used in friction modeling~\cite{Gnecco2001, fusco_velocity_2005, Brazda2018a,deWijn2012, norell2016emergent}, %\xc{which} % AS: this separates the subsject "potential" from the verb "involves"!
involves few nonzero Fourier components, resulting in few (or even no) non-zero terms in the sum of Eq.~\eqref{Uofr}.
%\xc{. This leads to a flat potential energy landscape}.
% AS: It is only flat if there is no term a part from V(0), but (usually) you have the term associated with the sinusoidal wave function.
%
%If all CLVs $\mathbf{\Omega}$ happen on null Fourier components, then the resulting interlocking potential is unrealistically flat.
%Note however that
In contrast however,
%the
our
example potential constructed as a sum of Gaussian wells,
%\xc{in our}
Eq.~\eqref{vperiodic},
involves infinitely many nonzero Fourier components, so that all terms in  Eq.~\eqref{Uofr} are potentially relevant.
%\xc{XIN: I feel that this paragraph should appear later, when we discuss the figure 1(d,h,m). Putting it here is a bit distractive and hard to understand.
%AS: I think it fits here where we introduce the general form of the potential energy. Then we refer to this part later. But introducing it in Sec IV would break the flow, I fear.
%NICK: let's leave it here. I understand Xin's concern, but it's something strictly related to eq. (4), so we have to tell something about this here...}

%The dependence of the potential $U(\mathbf{r}_\mathrm{c})$ upon the relative orientation of the two contacting lattices enters the expression in Eq. \eqref{Uofr} implicitly via the CLVs $\mathbf{\Omega}$ \xc{Xin: I suggest move this sentence to the place right after "and the summations run from ${-\infty}$ to ${\infty}$"}.
%
The focus of the present paper is how the shape of $U(\mathbf{r}_\mathrm{c})$ and the consequent friction phenomenology change (often dramatically), depending on the specifics of the intersection of the reciprocal lattices of the two contacting crystals.
%\xc{Xin: I suggest move this sentence to the place right after "so that all terms in  Eq.~\eqref{Uofr} are potentially relevant"AS: I do agree. The part about the angle dependence above breaks the flow. Like this works better?}.

\section{Algebraic basics}
\label{sec:lattice_algebra}
A lattice in $d$ dimension is the set of all vectors obtained as linear combinations with integer coefficients of a set of primitive vectors $\{\mathbf{R}_1,\mathbf{R}_2,...,\mathbf{R}_d\}$.
For the lattice $\mathcal{R}$ generated by this set of primitive vectors we use the following notation:
\begin{equation} \label{lattice}
\begin{split}
\mathcal{R} &= \left\{
\sum_{i}^{d} n_i\mathbf{R}_i : n_i \in \mathbb{Z}\right\} \\
&\stackrel{\text{def}}{=} [\mathbf{R}_1, \mathbf{R}_2,..., \mathbf{R}_d]
\,.
\end{split}
\end{equation}

The dual of an arbitrary set of vectors $\mathcal{R}$ (not necessarily a lattice), indicated by $\hat{\mathcal{R}}$, is defined as the set of all elements $\mathbf{Q} \in \mathbb{R}^d$, such that the scalar product $\langle \mathbf{Q}, \boldsymbol{\rho}\rangle$ is an integer multiple of $2\pi$ for all $\boldsymbol{\rho} \in \mathcal{R}$, i.e.
\begin{equation}\label{dual}
\hat{\mathcal{R}} =
\left\{\mathbf{Q} : \langle \mathbf{Q}, \boldsymbol{\rho}\rangle = 2\pi k,
k \in \mathbb{Z}, \forall \boldsymbol{\rho} \in \mathcal{R} \right\}
.
\end{equation}
Note that we include a factor $2\pi$, usually omitted from the standard mathematical definition of a dual.
With such convention,
%the dual of a lattice coincides with its reciprocal lattice
the dual of the lattice is its reciprocal lattice.
%\xc{Change to "With such convention, $\hat{\mathcal{R}}$ is its reciprocal lattice if $\mathcal{R}$ is a lattice"? AS: it is a general feature, I wouldn't refer to $\mathcal{R}$ specifically. But we can put it as an example indeed. Something like this?}

Let $\mathcal{S}$ be the lattice generated by a second set of primitive vectors $\{\mathbf{S}_1,\mathbf{S}_2,...,\mathbf{S}_d\}$.
The linear sum $\mathcal{S} + \mathcal{R}$ (defined as the set of all sums of one element in $\mathcal{S}$ plus another element in $\mathcal{R}$ %\xc{Xin: Since the linear sum appear in many figures, how about defining it by using an equation instead of a sentence? EMA: I think it is sufficiently clear like this. It is already quite heavy.})
of the two lattices is generally {\em not} a lattice, but it is a set of vectors, of which we wish to evaluate the dual.

It is a known properties of duals~\cite{micciancio2002complexity,Micciancio2010lattice_web} that $\hat{\mathcal{R}} \cap \hat{\mathcal{S}} = \widehat{\mathcal{S} +\mathcal{R}}$, where
%and \xc{Xin: Remove "and"?} AS: yes, looks out of place.
$\hat{\mathcal{R}} \cap \hat{\mathcal{S}}$ is the intersection of the individual duals $\hat{\mathcal{S}}$ and $ \hat{\mathcal{R}}$.

We adopt the following notation for the
$d=2$
problem %\xc{of contact friction}
at hand: the lattice $\mathcal{R}$ and its reciprocal
%\xc{Xin: a few lines above we call it "dual", I guess we change "reciprocal" here to "dual"? AS: a reciprocal (lattice) is always a dual (of a set), but not vice versa. Here, R and S are always lattices, so they have a reciprocal lattice. Instead O and Omega may not be lattice, so they have a dual but not a reciprocal. So, I would leave reciprocal here as we refer directly to R and S. It's good to use the lexicon familiar to cristallographers where we can. Xin: I see, I made some suggestions above about the relation between dual and reciprocal. Please see if it is appropriate.}
$\mathcal{Q} := \hat{\mathcal{R}}$ refer to the overlayer crystal;
$\mathcal{S}$ and its reciprocal $\mathcal{G} := \hat{\mathcal{S}}$ refer to the periodic substrate potential.
We also indicate the linear sum as $\mathcal{O} :=  \mathcal{S} +\mathcal{R}$ and its dual $\Omega :=  \hat{O} = \hat{\mathcal{R}} \cap \hat{\mathcal{S}}$.
By construction, $\Omega$ is then the set of
%coincidence vectors
CLVs in reciprocal space, namely those belonging to both reciprocal lattices $\hat{\mathcal{R}}$ and $\hat{\mathcal{S}}$.
For this reason, we will occasionally refer to $\Omega$ as the {\em coincidence set}.
%{\bf START}
%\xc{Xin: I rewrite the following paragraph:}
%Note that any arbitrary lattice $\mathcal{R}$ can be mapped to a linearly-transformed lattice $\mathcal{R}^{*}= [\mathbf{S}^{-1} \mathbf{R}]$. The explicit isomorphism is $\mathbf{S}^{-1}$ where $\mathbf{S}$ is the matrix whose columns are the vectors $\{\mathbf{S}_1,\mathbf{S}_2\}$ that generate $\mathcal{S}$. With this transformation, $\mathcal{S}^*=[\mathbf{S}^{-1} \mathbf{S}]=[\mathbb{1}]$, namely the hypercubic lattice of unit lattice spacing. Its dual (or reciprocal) is simply $\mathcal G^* = \widehat{\mathcal{S}^*}=2\pi[\mathbb{1}]$. In practice then, to classify the structure of $\mathcal{O}$, one can equivalently focus on the linear sum $\mathcal{O}^* = \mathcal{R}^*+[\mathbb{1}]$, i.e. the  $\mathbf{S}^{-1}$-transformed adsorbate lattice with a unit hypercubic lattice. The corresponding dual $\Omega^{*} =  \widehat{O^{*}} = \widehat{\mathcal{R^{*}}} \cap 2\pi[\mathbb{1}]$.
%For simplicity, all lattices and their linear sums are $\mathbf{S}^{-1}$-transformed in the rest of the paper. For compactness' sake we also omit all stars. Figure~\ref{fig:coverages} already adheres to this convention

Note that any two arbitrary %2D \xc{Xin: Since we also discuss the degree of incommensurability in 1D in the next section, I guess we don't need to refer to "2D" here and in the next paragraph, since these lattice theories apply to any dimension.
%AS: fair point, in the end is just a change of basis set. I think we can drop the 2D, let's see if Emanuele agrees. Xin: if agree, we need to change the "unit square lattice" to "unit hypercubic lattice" ?}
lattices $\mathcal{R}$ %\as{\st{in contact with another arbitrary 2D lattice}
and $\mathcal{S}$ can be mapped to a linearly-transformed lattice $\mathcal{R}^*$ %\as{\st{in contact with a}
and a unit square lattice $[\mathbb{1}]$.
%\xc{Xin: change to “Note that the linear sum of any arbitrary 2D lattice $\mathcal{R}$ and $\mathcal{S}$ can be mapped to (i.e. isomorphic to) the linear sum of a linearly-transformed lattice $\mathcal{R}^*$ and a unit square lattice $[\mathbb{1}]$.”?
%AS: I am not understanding here, we do not map the linear sum, but the two distinct lattices. In other words, the pair of lattices $(R,S)$ is isomorphic to the pair $(R^*, 1)$ (via the isomorphism induces by $S^-1$). Right?
%Xin: What is a "pair" of lattices? How they "pair" each other? You need to define what is the outcome of "R in contact with S" if you want to map one contact to another.
%AS: a pair as in literally two element. But I think I see why it seems unclear. We say "in contact with" but it actually this isomorphism has nothing to do with the contact. We are just describing the lattices in the "coordinate of S", i.e. in the coordinates where S is the unit square. Better?
%}
The explicit isomorphism is $\mathbf{S}^{-1}$ where $\mathbf{S}$ is the matrix whose columns are the vectors $\{\mathbf{S}_1,\mathbf{S}_2\}$ generating
%\xc{Xin: change to "that generates"? AS: the -ing form like should be exactly equivalent to what you're suggesting, no?}
$\mathcal{S}$.
With this transformation, $\mathcal{S}^*=[\mathbb{1}]=[\mathbf{e}_x,\mathbf{e}_y]$, namely
%the unit square lattice
the square lattice of unit lattice spacing% AS: maybe the first time we spell it out as clearly as possible.
, and $\mathcal G^* \equiv \widehat{\mathcal{S}^*}$ consists of all reciprocal vectors whose Cartesian components are $2\pi \times$integers.
In practice then, to classify the structure of $\mathcal{O}$, one can equivalently focus on
%the linear sum
$\mathcal{O}^* = \mathcal{S}^*+\mathcal{R}^*$,
%\xc{Xin: if you can equivalently focus on $\mathcal{O}^*$, then it means $\mathcal{O}^*$ and $\mathcal{O}$ are isomorphic to one another? AS: yes, as O is the sum of two isomophic elements and the isomorphism is a linear map, I think.}
%of the
i.e. the linear sum of the
unit square lattice with a properly  transformed lattice $\mathcal{R}^*=[\mathbf{R}^*_1,\mathbf{R}^*_2] = [\mathbf{S}^{-1} \mathbf{R}]$. %The corresponding dual is $\Omega^{*} (=  \widehat{O^{*}} ) = \widehat{\mathcal{R^{*}}} \cap \widehat{\mathcal{S^{*}}}$.

In the following we stick to the $\mathbf{S}^{-1}$-transformed lattices $\mathcal{R}^*$ and $\mathcal{S}^*$.
For compactness' sake we omit all stars.
Figure~\ref{fig:coverages} already adheres to this convention.

In the next section, we show that %\xc{, for the 2D problem at hand,}
three different categories of $\mathcal O$ (or its dual $\Omega$) can arise, depending on the mutual (in)commensuration of $\mathcal{R}$ and $\mathcal{S}$.
We further show that the specific type of $\mathcal O$ pertinent to a given interface between two flat crystalline surfaces bears important implications for the friction exhibited by this interface when the surfaces are set in relative motion.

\section{Three categories of commensurability}
\label{sec:infinite_size}
For a 1D interface, the degree of commensuration is dictated by the ratio $a/b$ between the lattice spacing of the adsorbate and
%the one
that of the substrate~\cite{frenkel1938model,aubry1983FKanal,braun2004frenkel}:
if this ratio is a rational number, then the two lattices are commensurate; if this ratio is irrational, the lattices are incommensurate.
In the rational case the coincidence set $\Omega$ %= 2\pi(a/b)[\mathbb{1}] \cap 2\pi[\mathbb{1}]$}
is a 1D lattice where the periodicity is given by
the denominator of the ratio.
%the largest common denominator between the spacings. EMA: WRONG! NICK: Well spotted!!
Then the linear-sum space $\mathcal{O}$ % = (a/b)[\mathbb{1}]+[\mathbb{1}]$}
is also
a discrete set of points repeating along the 1D line, i.e. a 1D lattice. % \xc{Xin: change "a discrete set of points  repeating along the 1D line" to "also a lattice"? Since a dense coverage as described below is also a discrete set of points even though the distance between the points can be infinitely small...}.
In the irrational case there
are no CLVs
%\xc{is no coincidence set in reciprocal space}
and $\Omega$ contains the null element only.
Hence %, there is no periodicity and
the sum space $\mathcal{O}$ %\xc{in real space}
covers densely the 1D line. %\xc{, i.e. for arbitrary vector $\mathbf{x}$ in the 1D line, there is always a $\mathbf{R}\in\mathcal{O}$ such that their distance $|\mathbf{R}-\mathbf{x}|$ can be infinitely small}
For a rigid contact, the rational case has finite friction, while the irrational case is frictionless.

In 2D the linear sum $\mathcal{O}$ is a set of vectors in $\mathbb{R}^2$.
The friction experienced by the overlayer when it is translated relative to the substrate potential is determined by the geometric mutual commensuration relations of the vectors in  $\mathcal{R}$ and $\mathcal{S}$, and their implications for $\mathcal{O}$. %\xc{Xin: I found this sentence rather distractive here, a similar sentence "The specific type of $\Omega$ pertinent ..." also appears in the last paragraph of previous section. Maybe remove this one here.}.
These relations characterize how the real plane $\mathbb{R}^2$ is covered by the vectors in the sum space $\mathcal{O}$.
This coverage occurs in one of three qualitatively different types:

\begin{itemize}
\item[(A)] a sparse coverage by a discrete set of points forming a 2D lattice -- Fig.~\ref{fig:coverages}c,
\item[(B)] a ``comb'' coverage by an array of parallel lines -- Fig.~\ref{fig:coverages}g, and
\item[(C)] a dense \footnote{A set $\mathcal{O}$ is dense if it has the following property: for any $\epsilon >0 \in \mathbb{R}$ and for any point $\mathbf P$ in the vector space, there exists a $\mathbf{O}_1 \in \mathcal{O}$ such that  $|\mathbf{P}-\mathbf{O}_1|<\epsilon$} area coverage -- Fig.~\ref{fig:coverages}k.
\end{itemize}
These geometric alternatives result in three dramatically different patterns of interlocking potential, illustrated in the rightmost column of Fig.~\ref{fig:coverages}, and therefore different friction properties.
In the following subsections we detail the conditions determining these coverage categories in terms of the reciprocal lattice  $\mathcal{Q} = \hat{\mathcal{R}}$ %of the ($\mathbf{S}^{-1}$-transformed) lattice $\mathcal{R}$,
and of the coincidence set $\Omega$.

\begin{table*}[]
    \centering
    \begin{tabular}{c|c|c}
         Type & Condition on the $\mathcal{Q}=\hat{\mathcal R}$ lattice &
coverage $\mathcal{O}$
\\
         \hline
         A & $\exists \mathbf{Q}_1 , \mathbf{Q}_2\in \mathcal{Q}\setminus\{\mathbf{0}\}: Q_{i\mu}/(2\pi) \in \mathbb{Z} \text{ and }
        \nexists \alpha\in \mathbb{R}: \mathbf{Q}_1 = \alpha \mathbf{Q}_2$
        & 2D discrete \\
         B &
         $\exists \mathbf{Q}_1\in \mathcal{Q}\setminus\{\mathbf{0}\}: Q_{1\mu}/(2\pi) \in \mathbb{Z}$ and $\forall\mathbf{Q}_2\in \mathcal{Q}: Q_{2\mu}/(2\pi) \in \mathbb{Z}$,
         $\exists \alpha\in \mathbb{R}: \mathbf{Q}_2 = \alpha \mathbf{Q}_1$
         & 1D discrete   $\otimes$ 1D dense \\
         C & $\nexists \mathbf{Q}\in \mathcal{Q}\setminus\{\mathbf{0}\}:
         Q_{\mu}/(2\pi) \in \mathbb{Z}
         $
         & 2D dense
    \end{tabular}
    \caption{The three categories of commensurability, based on how the reciprocal lattice vectors $\mathbf{Q}\in\mathcal{Q}=\hat{\mathcal R}$ of the overlayer crystal match the reciprocal $\mathcal{G} = \hat{\mathcal S}$ of the substrate unit square lattice.
    }
    \label{tab:conditions}
\end{table*}

\subsection{Discrete coverage}
Discrete coverage occurs when there exist two linearly independent vectors $\{\mathbf{Q}_1, \mathbf{Q}_2\}$ % \xc{Xin: I would avoid using $\{\mathbf{Q}_1, \mathbf{Q}_2\}$ here since they would be regarded as primitive vectors of the $\mathcal{Q}$ space following our previous convention. Maybe change them to $\{\mathbf{Q}_\alpha, \mathbf{Q}_\beta\}$?}
in the dual $\mathcal{Q}$ of the adsorbate crystal lattice $\mathcal{R}$ such that all components of these vectors are $2\pi\times $ integer numbers, i.e. $Q_{i\mu}/(2\pi) \in \mathbb{Z}$ with $i=1,2$ and $\mu=x,y$.
%
%Since $\mathcal G =2\pi[\mathbb{1}]$,
The above condition means that $\mathbf{Q}_1, \mathbf{Q}_2$ exist in $\mathcal G $ as well, i.e. $\mathbf{Q}_1, \mathbf{Q}_2\in\Omega=\mathcal{Q} \cap \mathcal{G}$.
%%This
%condition amounts to
%%the coincidence of finding
%\as{the existence of}
%$\mathbf{Q}_1$ and $\mathbf{Q}_2$
%also in the dual $\mathcal{G}$ of the substrate lattice.
%\textcolor{red}{ERIO: "the coincidence of finding" ...what does that mean? The word coincidence seems improperly used... You mean to say "the existence"? AS: I guess we wanted to stress that it's not mandatory for this to happen, but you're right, call it a coincidence is inappropiate.}
In this case, summarised in the first row of Table~\ref{tab:conditions}, $\Omega$
%$\Omega= \hat{O} = \hat{\mathcal{R}} \cap \hat{\mathcal{S}}$
is a 2D lattice.
%The corresponding linear sum $\mathcal{O}$ is also a 2D lattice %consists of a discrete set of points
%as in the example of Fig.~\ref{fig:coverages}c.
%\as{The corresponding linear sum $\mathcal{O}$ is also a 2D lattice %consists of a discrete set of points
%as in the example of Fig.~\ref{fig:coverages}c.
%}
As a corollary of the above condition, all reciprocal vectors $\mathbf{Q}_i\in \mathcal{Q}$ are such that their components $Q_{i\mu}/(2\pi)$ are rational numbers; see Methods Section \ref{sec:corollary_rational} for a proof.

%In terms of real space, this kind of discrete coverage occurs when all elements in ${\mathcal R}$ have both components $\text{R}_{j,\mu} \in \mathbb{Q}$ \xc{Xin: this Q-like symbol means the set of all rational numbers?}.
%
%Under these conditions, the sum space $\mathcal{O}$ consists of a discrete set of points, as in the example of Fig.~\ref{fig:coverages}c.
We pick two primitive vectors $\mathbf{\Omega}_1$ and $\mathbf{\Omega}_2$ of this coincidence reciprocal lattice $\Omega = [{\mathbf \Omega}_1,{\mathbf \Omega}_2]$, as illustrated in Fig.~\ref{fig:coverages}b. %\xc{Xin: how about move this sentence to the beginning of next paragraph?}.
The resulting interlocking potential, as expressed in Eq.~\eqref{Uofr}, exhibits a non-vanishing corrugation along both independent directions of ${\mathbf \Omega}_1$ and ${\mathbf \Omega}_2$. %\xc{Xin: I would remove this sentence}.
The interlocking potential energy landscape for the center-of-mass translation takes on a nontrivial shape, such as e.g.\ the one depicted in Fig.~\ref{fig:coverages}d. It can be expressed as
\begin{align}
\label{Uofr_discrete}
U(\mathbf{r}_\mathrm{c}) &=
\sum_{j_1, j_2} \tilde{V}_\mathcal{S}(j_1 \mathbf{\Omega}_1 + j_2 \mathbf{\Omega}_2 ) \, \nonumber \\
&\quad \times \text{exp}\left( i (j_1
\mathbf{\Omega}_1 + j_2 \mathbf{\Omega}_2 )
\cdot
\mathbf{r}_\mathrm{c} \right)
\,,
\end{align}
where the sum runs over all integeres $j_1, j_2 \in \mathbb{Z}$.
Fig.~\ref{fig:coverages}d reports this function as resulting from the Fourier components of the potential of Eq.~\eqref{vperiodic}.
The potential energy landscape reveals a new lattice periodicity, precisely the one emerging in the sum space $\mathcal{O}$, Fig.~\ref{fig:coverages}c. %,  whose spatial
In general, this new periodicity is either equal %(\xc{Xin: ? I think in Figure 3c it should always be "shorter than", is that right?})
to or shorter than that of the original substrate.
% NICK: perhaps we could find a better word for "smaller". What is smaller is the period, no the periodicity, but in general there is no single period, so....

When the monolayer is forced across the periodic surface, it will move more easily along certain directions than along others, due to potential-barrier directional anisotropies.
%For
In the example of Fig.~\ref{fig:coverages}d, the easy directions are the %corrugation-lattice
substrate lattice
symmetry directions which avoid the maxima of the energy landscape.

Motion along these preferential directions is usually named {\em directional locking}.
This phenomenon was observed experimentally in several systems,
e.g., for a triangular colloidal crystalline 2D cluster %made slide
sliding
across a triangular substrate~\cite{Cao2019a,Cao2021,bohlein2012experimental,korda2002kinetically},
for AFM-pushed gold islands on MoS$_2$~\cite{Trillitzsch2018},
and for magnetic vortex lattice under a Lorentz force~\cite{villegas2003directional}, % AS: is this the right way to refer to these?
% COPY REFERENCES FROM NATURE COMMUNICATIONS | (2020)11:4657 https://doi.org/10.1038/s41467-020-18429-1 works cited at pages 6-8.
% AS: I copied all the directional references relevant. I added a simulations part for the Reichhardts.
and computationally in different contexts~\cite{reichhardt1999phase,gopinathan2004statistically,frechette2009directional,speer2010exploiting,reichhardt2011dynamical,pelton2004transport}.

\subsection{Line coverage}
A second nontrivial condition occurs when the coverage $\mathcal{O}$ is neither discrete nor dense in the whole space.
This  condition is realized when the dual
%space % AS: why space? It is a set
set
$\Omega = \mathcal{Q} \cap \mathcal{G}$ is a 1-dimensional lattice, as in Fig.~\ref{fig:coverages}f.
In practice, one can identify a nonzero reciprocal lattice vector $\mathbf{Q}_1 \in \mathcal{Q}$ with the property of having both integer components $Q_{1\mu}/(2\pi) \in \mathbb{Z}$.
All other vectors in $\mathcal{Q}$ either are multiples of $\mathbf{Q}_1$ or have at least a component divided by $2\pi$ that is an irrational number.
Pick one of the two shortest nonzero inversion-symmetric CLV, namely the vectors with the properties of $\mathbf{Q}_1$, and call it $\mathbf{\Omega}$.
As a result %of
the coincidence set $\Omega$ contains a unique linearly-independent direction, that of $\mathbf{\Omega}$.
In other words, $\Omega$ is a 1-dimensional lattice $\Omega = [\mathbf{\Omega}]$, simply the set of all integer multiples of $\mathbf{\Omega}$.
When this condition is met, the linear sum space $\mathcal{O}$ is a set of  discretely-spaced parallel lines aligned perpendicularly to $\mathbf{\Omega}$, and covered densely in a 1-dimensional sense, see Fig.~\ref{fig:coverages}g.
This condition is summarised in the second row of Table~\ref{tab:conditions}.

In terms of the shortest CLV $\mathbf{\Omega}$, the
% substrate average potential % AS: stick to one name.
interlocking potential
energy of Eq.~\eqref{Uofr}
becomes:
\begin{equation}
\label{Uofr_linear}
U(\mathbf{r}_\mathrm{c}) =
\sum_{n} \tilde{V}_\mathcal{S}(n\mathbf{\Omega}) \, \text{exp}(i n\mathbf{\Omega}\cdot \mathbf{r}_\mathrm{c})
\,,
\end{equation}
which is the analog of Eq.~\eqref{Uofr_discrete} except now the sum spans the 1D lattice generated by $\mathbf{\Omega}$.
As a consequence, the interlocking potential is a function uniquely of the component of the displacement vector $\mathbf{r}_\mathrm{c}$ in the $\mathbf{\Omega}$ direction.
Explicitly, and importantly, $U(\mathbf{r}_\mathrm{c})$ is completely independent of the displacement component of $\mathbf{r}_\mathrm{c}$ perpendicular to $\mathbf{\Omega}$.
The function $U(\mathbf{r}_\mathrm{c})$ can be pictured as a periodic set of parallel straight troughs separated by straight hill ridges, see Fig.~\ref{fig:coverages}h.
Troughs and ridges are aligned perpendicular to $\mathbf{\Omega}$.
As a result, the contact behaves ``as commensurate'', and exhibits a finite static friction, in all directions, except for this direction perpendicular to $\mathbf{\Omega}$.
In this direction it behaves ``as incommensurate'' with {\em vanishing} static friction, since the through bottom is ``flat'', i.e.\ has a constant energy.

A natural name for this condition is {\em directional structural lubricity}.

We are not aware of any previous work where this regime of frictionless nanohighways is hypothesized, nor any existing experiment where this condition was pointed out.
In Sec.~\ref{sec:approxCLV} below we  report its realization in a colloidal experiment, and propose possible heterogeneous contacts between 2D nanomaterials where it should also be accessible.

We realize that in general, if two lattices share the same rotational symmetry of order $n > 2$, then there arise either two or no linearly independent CLVs in $\Omega$.
Therefore  the resulting contact cannot belong to this type B.
In Methods Section~\ref{sec:rotation_arg} we report an argument to illustrate this point.

\subsection{Dense coverage}
The final possibility occurs when no nonzero reciprocal lattice vector in $\mathcal{Q}$ exists
such that both its components divided by $2\pi$ are integer.
This statements amount to say that the coincidence set is $\Omega = \{{\mathbf 0}\}$, as summarised in the final row of Table~\ref{tab:conditions}.
The corresponding sum
%space % AS: not a space, just a set
set
$\mathcal{O}$ is dense, as shown in Fig.~\ref{fig:coverages}k.

The Fourier expansion \eqref{Uofr} only includes the null vector: as a result, the
%substrate average potential
interlocking potential energy
is perfectly flat
\begin{equation}
\label{Uofr_dense}
U(\mathbf{r}_\mathrm{c}) = \tilde{V}_\mathcal{S}(\mathbf{0})
\,,
\end{equation}
as in Fig.~\ref{fig:coverages}m.
No energy is gained or spent in rigidly translating the monolayer by an arbitrary amount in an arbitrary direction.
Static friction vanishes.
This is the well-known condition of structural superlubricity of fully incommensurate lattices~\cite{Vanossi2020a,dienwiebel2004,Song2018RobustSuperlub}.

Interestingly, two lattices can even share CLVs in \textit{real} space, but still have a dense sum space $\mathcal{O}$.
This occurs in the example of Fig.~\ref{fig:coverages}i--m:
these two lattices happen to have infinitely many coincidence points along the substrate direction $\mathbf{S}'=(2,1)$, i.e.\ the intersection $\mathcal{R} \cap  \mathcal{S} = [\mathbf{S}']$
(orange-edged blue circles in Fig.~\ref{fig:coverages}i).
However this real-space intersection is irrelevant to the lubricity of this contact.
Instead, the relevant coincidence set $\Omega$ contains only the null vector (Fig.~\ref{fig:coverages}j), and, as a result, the 2D linear space is covered densely (Fig.~\ref{fig:coverages}k), the average potential energy is flat (Eq.~\eqref{Uofr_dense} and Fig.~\ref{fig:coverages}m), and the static friction vanishes equally in  all directions.

\section{Finite size}\label{sec:finite_size}
The formulas \eqref{Uofr_discrete}-\eqref{Uofr_dense} for the
%substrate average potential energy
interlocking potential
$U(\mathbf{r})$ are valid for an infinite %adsorbate
2D crystalline overlayer
interacting with an infinite periodic surface.
In practice, however, no contact is infinitely extended.
In real-life conditions the overlayer region in contact with the substrate forms a crystalline cluster of finite size $N$.
Its finite size and shape do affect the details of the
interlocking
potential energy $U(\mathbf{r}_\mathrm{c})$:
in the following,
we derive analytical expressions for finite-size contacts and their implications on the frictional behaviour.

%Like
As
in Eq.~\eqref{UofrDef}, we express  the position of each particle as a function of the cluster center of mass displacement $\mathbf{r}_\mathrm{c}$ and lattice vectors: $\mathbf{r}_j=\mathbf{r}_\mathrm{c}+\mathbf{R}_j$, where precisely the list of the $N$ translations $\mathbf{R}_j\in {\mathcal R}$ defines the cluster shape and size.

Starting from Eq.~\eqref{UofrDef}, we write the explicit expression for the interlocking potential energy $U(\mathbf{r}_\textrm{c})$ in terms of the substrate-potential Fourier decomposition.
\begin{align}
%\begin{split}
% AS: This is EXACTLY eq 2, do we really need to write it again?
%U(\mathbf{r}_\textrm{c}) =& \frac 1N \sum_j V_\mathcal{S}(\mathbf{R}_j + \mathbf{r}_\textrm{c})	 \nonumber \\
U(\mathbf{r}_\textrm{c}) =&
\sum_{\mathbf{G} \in
\mathcal{G}
}
\tilde{V}_\mathcal{S}(\mathbf{G}) \exp(i\mathbf{G}\cdot\mathbf{r}_\textrm{c})
 \nonumber \\
& \times
\frac 1N \sum_j \exp(i\mathbf{G}\cdot\mathbf{R}_j)
 \label{eqUrN1} \\
=&
\sum_{\mathbf{G} \in \mathcal{G}} \tilde{V}_\mathcal{S}(\mathbf{G}) \exp(i\mathbf{G}\cdot\mathbf{r}_\textrm{c})
\nonumber \\
& \times
\frac 1N \sum_j \exp(i( \mathbf{G} - \mathbf{Q}) \cdot\mathbf{R}_j)
\,,
\label{eqUrN2}
%\end{split}
\end{align}
where $\mathbf{Q}\in \mathcal{Q}$ is an arbitrary vector of the reciprocal lattice of the adsorbate.

In going from Eq.~\eqref{eqUrN1} to Eq.~\eqref{eqUrN2} we use the property
of the reciprocal vectors $\mathbf{Q}$ that $\exp(i\mathbf{Q}\cdot \mathbf{R}_j) =1\, \forall j$.
We take advantage of the freedom in the choice of $\mathbf{Q}\in \mathcal{Q}$ to pick $\mathbf{Q} = \bar{\mathbf Q}$ such that $|\mathbf{G}-\bar{\mathbf{Q}}|$ is minimum, and we define
\begin{equation}\label{deltaOmega}
    \delta\mathbf{\Omega}(\mathbf{G}) = \mathbf{G}-\bar{\mathbf Q}
\end{equation}
to ensure that $\delta\mathbf{\Omega}(\mathbf{G})$ fits in the Wigner-Seitz cell of the adsorbate reciprocal lattice, i.e.\ in the first Brillouin zone of $\mathcal{R}$.
Note how this notation relates with the infinite-size classification of Sec.~\ref{sec:lattice_algebra}: if $\mathbf{G}$ is a common vector between the substrate and adsorbate reciprocal lattices, i.e. $\mathbf{G} \in \Omega$, then $\delta \mathbf{\Omega}(\mathbf{G}) = \mathbf{0}$ for that $\mathbf{G}$.

The
%average
interlocking
potential energy in Eq.~\eqref{eqUrN2} now reads
\begin{align}
U(\mathbf{r}_\textrm{c}) =&
\sum_{\mathbf{G} \in \mathcal{G}} \tilde{V}_\mathcal{S}(\mathbf{G}) \exp(i\mathbf{G}\cdot\mathbf{r}_\textrm{c})
\nonumber \\
& \times
\frac 1N \sum_j \exp(i\delta\mathbf{\Omega}(\mathbf{G})\cdot\mathbf{R}_j)
 \label{eqUrN4} \\
=&
\sum_{\mathbf{G} \in \mathcal{G}} \tilde{V}_\mathcal{S}(\mathbf{G}) \exp(i\mathbf{G}\cdot\mathbf{r}_\textrm{c}) \, W(\delta\mathbf{\Omega}(\mathbf{G}), N)
\,,
\label{eqUrN5}
\end{align}
where we defined the weight factor
\begin{equation}
    \label{eq:Wdef}
W(\delta \mathbf{\Omega}(\mathbf{G}), N)= \frac 1N \sum_j \exp(i \delta\mathbf{\Omega}(\mathbf{G}) \cdot \mathbf{R}_j)
\,.
\end{equation}
In the following, we omit the explicit dependence of $\delta\mathbf{\Omega}(\mathbf{G})$ on the substrate lattice vector $\mathbf{G}$, $\delta\mathbf{\Omega}=\delta\mathbf{\Omega}(\mathbf{G})$.
%, or specify it as a subscript to resolve ambiguous cases, $W_\mathbf{G}:=W(\delta\mathbf{\Omega}(\mathbf{G}), N)$. EMA: Never actually used!!!

Equation~\eqref{eqUrN5} expresses the interlocking
% substrate average
potential energy $U(\mathbf{r}_\textrm{c})$ as a Fourier summation over all the components $\tilde{V}_\mathcal{S}(\mathbf{G})$ of the substrate potential \cite{Cao2019a}.
The novelty of finite size compared to Eq.~\eqref{Uofr} is that the summation of Eq.~\eqref{eqUrN5} is not limited to CLVs and it involves the extra size-dependent weight factor defined in Eq.~\eqref{eq:Wdef}.

In the $N\to \infty$ limit, this weight $W(\delta \mathbf{\Omega}, N)$ vanishes for those Fourier components with $\delta \mathbf{\Omega}\neq \mathbf 0$.
Therefore, only the $\delta \mathbf{\Omega} = \mathbf{0}$ components determine the overall corrugation of the infinite-size crystal, in agreement with the classification discussed in the previous Section.

% NICK: the following sentence is a bit of a repetition of concepts already expressed in the previous paragraph:
At finite size, $W(\delta \mathbf{\Omega}, N)$ needs not vanish for any
${\mathbf{G}}$, regardless of whether
$\delta \mathbf{\Omega}$ vanishes or not.
As a consequence, a priori any Fourier component $\tilde{V}_\mathcal{S}({\mathbf{G}})$ of the substrate potential can contribute to the interlocking potential energy $U(\mathbf{r}_\textrm{c})$.

In general, the weight $W(\delta \mathbf{\Omega}, N)$ is a nontrivial function of $\delta \mathbf{\Omega}$, which depends on the cluster shape.
However, for special shapes, analytic expressions for $W(\delta \mathbf{\Omega}, N)$ can be derived.

\subsection{Special shapes}
\label{specialshapes:sec}
As a concrete example,
consider a  parallelogram-shaped
cluster of $N$ particles whose particle positions can be expressed as  $\mathbf{R}_j = j_1 \mathbf{R}_1 + j_2 \mathbf{R}_2 $, with integer $j_1,j_2 = -(\sqrt{N}-1)/2, \dots, (\sqrt{N}-1$)/2, where $\sqrt{N}$ is assumed to be an odd integer, and $\mathcal{R}=[\mathbf{R}_1, \mathbf{R}_2]$.
By construction, the cluster center of mass coincides with the particle indexed by $j_1=j_2=0$.
For this cluster shape, the weight function of Eq.~\eqref{eq:Wdef} can be written as
\begin{align}\nonumber
 W(\delta \mathbf{\Omega}, N) &=  \frac 1N \sum_{j_1,j_2} \exp(i\delta \mathbf{\Omega}\cdot j_1\mathbf{R}_1) \exp(i\delta \mathbf{\Omega}\cdot j_2\mathbf{R}_2)
 \\\label{eq:Wdefanal}
	   & = \frac{1}{N}
	   \frac{\sin(\sqrt{N} \delta \mathbf{\Omega}\cdot \mathbf{R}_1 / 2) }{ \sin( \delta \mathbf{\Omega}\cdot \mathbf{R}_1 / 2)}
	   %\,
	   \frac{ \sin(\sqrt{N} \delta \mathbf{\Omega}\cdot \mathbf{R}_2 / 2) }{ \sin( \delta \mathbf{\Omega}\cdot \mathbf{R}_2 / 2)}
\,.
\end{align}
Each factor $f(x) = \sin(\sqrt{N}x/2)/\sin(x/2)$ in Eq.~\eqref{eq:Wdefanal} relates to the Fraunhofer diffraction from a narrow-slit grating~\cite{born2013principles}.
As a function of $x=\delta \mathbf{\Omega}\cdot \mathbf{R}_i$, each oscillating factor $f(x)$ peaks at $x=2n\pi$ (for $n\in\mathbb{Z}$), where it reaches its extreme values $(-1)^n \sqrt{N}$, i.e. $\pm \sqrt{N}$.
As a special consequence, whenever $\delta \mathbf{\Omega}$ vanishes, both fractions becomes equal to $\sqrt{N}$, leading to $W(\delta \mathbf{\Omega}, N)=1$: the  weight of the corresponding Fourier components in Eq.~\eqref{eqUrN5} is independent of size.

The peak width of $f(x)$ is inversely proportional to $\sqrt{N}$, and away from the peaks, say in the intervals $ 2\pi (n + N^{-1/2})< x < 2\pi (n+1-N^{-1/2})$ $f(x)$, oscillates around 0, with values of order 1.
As a result, for large cluster size, weights associated to both nonzero  $\delta \mathbf{\Omega}\cdot \mathbf{R}_1$ and $\delta \mathbf{\Omega}\cdot \mathbf{R}_2$ decay as $N^{-1}$.
Instead, when just one of these factors vanishes, we expect a nontrivial leading large-size behavior of the associated weight factor, typically as $N^{-1/2}$.
These observations account for the leading importance of the Fourier component in the
%average corrugation,
interlocking potential $U(\mathbf{r}_\mathrm{c})$
as detailed in the following subsections.

\subsection{Coincidence lattice vectors (CLVs)}
\begin{figure*}[tb]
    \centering
    \includegraphics[width=0.9\textwidth]{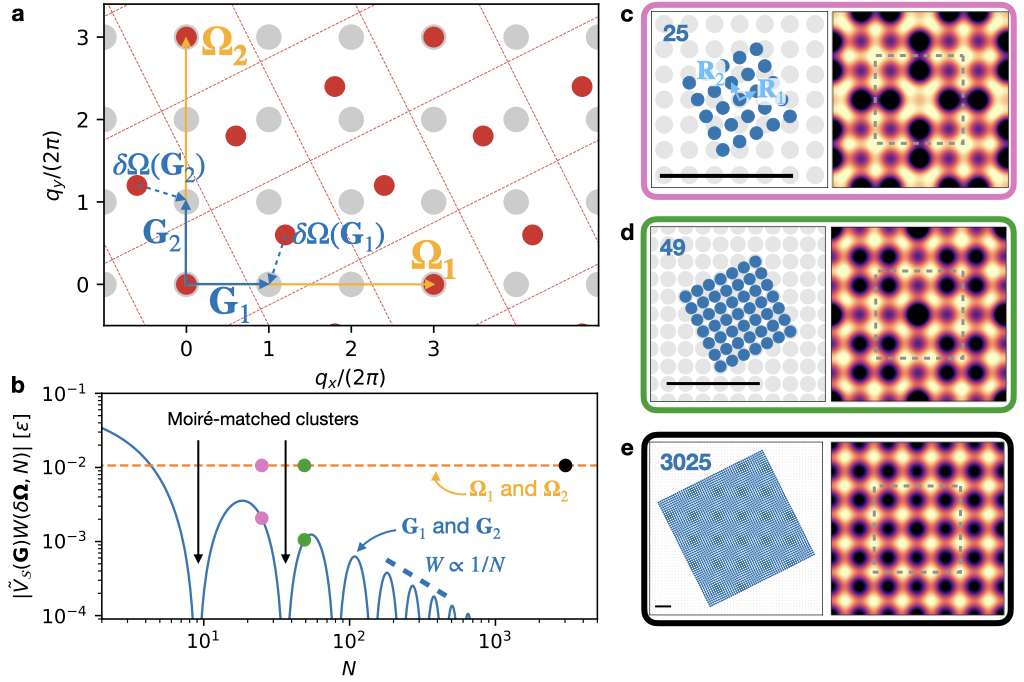}
    \caption{
    Finite size effects in a contact with discrete coverage.
    %\xc{Xin: I guess this part is mainly to show that, as cluster size increases, the non-CLV components on the potential energy landscape become disappear? If true, I guess we change the title of this figure accordingly. By reading the current title, I would wonder "is there no finite-size analysis for such commensurate contact before?". Why it is so important to be the 2nd figure? I suggest something like "\textbf{The role of coincidence lattice vectors in determining the potential energy landscape.}" or "\textbf{The relation between the coincidence lattice vectors and the potential energy landscape.}"}
    Finite-size analysis for the discrete-coverage (type A) condition in Fig.~\ref{fig:coverages}a--d, i.e.\ a square-lattice adsorbate with $|a| = \sqrt{5}/3$, and orientation $\theta_\mathrm{o} = \tan^{-1}(1/2)$.
    (a) Reciprocal lattices of the adsorbate ($\mathbf Q$ points, red dots) and of the substrate ($\mathbf G$ points, gray dots); orange arrows highlight two CLVs $\mathbf{\Omega}_{i=1,2}$; blue arrows indicate the substrate primitive vectors $\mathbf{G}_i$.
    Red dashed squares show the first Brillouin zone of the $\mathcal{Q}$ lattice at each lattice point $\mathbf Q$.
    The dashed blue arrows indicate $\mathbf{\delta\Omega}(\mathbf{G}_i)$, defined in Eq.~\eqref{deltaOmega}, i.e. the $\bar{\mathbf Q}$-translated  vectors $\mathbf{G}_{i=1,2}$.
    Note that $\mathbf{\delta\Omega}(\mathbf{\Omega}_i)=\mathbf 0$.
    (b) The size dependence of the Fourier components $\tilde{V}_\mathcal{S}W$ (Eq.~\eqref{eqUrN5}) associated with the CLVs $\mathbf{\Omega}_i$ (dashed orange line) and for the primitive vectors $\mathbf{G}_i$ (solid blue line).
    The pink, green and black circles mark the cluster sizes $N=25$, $49$, $3025$ of panels (c), (d), and (e), respectively.
    (c--e)
    %Average potential energy landscape
    Interlocking potential energy
    as a function of the cluster center-of-mass position $\mathbf{r}_\mathrm{c}$ across the square $[-1,1]\times[-1,1]$ for the displayed clusters.
    Solid black scalebars in panels (c--e) stand for five lattice spacings.
    The color scale is the same as in Fig.~\ref{fig:coverages}.
    The gray dashed line highlights the Wigner-Seitz cell of the $\mathcal S$ lattice, namely the area shown in Fig.~\ref{fig:coverages}d.
    }
    \label{fig:size_match}
\end{figure*}
When there are CLVs in the dual space, an infinite subset of $\mathcal{G}$ vectors (those belonging to $\Omega$) is associated to vanishing
$\delta \mathbf{\Omega}$.
For all these terms the weights $W(\delta \mathbf{\Omega}, N)$ equal unity, and as a result, the corresponding Fourier components in Eq.~\eqref{eqUrN5} contribute as much as for the infinite-size layer, independently of size.
All other Fourier components, characterized by nonzero $\delta \mathbf{\Omega}$, can contribute with
size-dependent weights.
This distinction between size-independent and size-dependent Fourier weights applies to
systems with discrete-$\mathcal{O}$ (type-A) and line coverage (type-B) geometries, as discussed for the infinite size-limit in Sec.~\ref{sec:infinite_size}.

Let us focus first on type-A contacts, where two linearly independent CLVs $\mathbf{\Omega}_1, \mathbf{\Omega}_2$, with $\delta\mathbf{\Omega}=\mathbf{0}$, exist.
Consider also two non-CLV vectors $\mathbf{G}_1,\mathbf{G}_2$, with $\delta\mathbf{\Omega}\neq\mathbf{0}$, as exemplified in Fig.~\ref{fig:size_match}a.
It is possible to obtain instructive results for the special parallelogram-shaped clusters introduced in previous section.
For $\mathbf{\Omega}_{1,2}$ and $\mathbf{G}_{1,2}$, Fig.~\ref{fig:size_match}b reports the Fourier amplitudes $\tilde{V}_\mathcal{S}$, modulated by the weights $W$,  computed according to Eq.~\eqref{eq:Wdefanal}.
The weight factor $W$ is identically unity at any size for the CLVs (orange dashed line), while it oscillates and decreases as a function of size for non CLVs (blue solid line).

To elucidate the origin of the $\mathbf{G}_{1,2}$-weight oscillations it is convenient to focus on a subset of cluster sizes.
Consider the clusters constructed as $\sqrt{N'}\times \sqrt{N'}$ repetitions of a parallelogram supercell constructed on a pair of independent real-space CLVs $\mathbf{R}^\text{CLV}_1$ and $\mathbf{R}^\text{CLV}_2$, namely vectors with all integer components  (e.g.\ the blue orange-edge dots in Fig.~\ref{fig:coverages}a).
The existence of these real-space CLVs is demonstrated in Methods Section~\ref{sec:corollary_rational}.
The supercell defined by these (usually non-primitive) vectors contains $K$ vectors and thus the number of particles is $N=KN'$.
By construction, these special clusters consist of an integer number of identical moir\'e tiles of $K$ particles in which the relative position of adsorbate and substrate is the same in all repeated units, since $\mathbf{R}^\text{CLV}_1$ and $\mathbf{R}^\text{CLV}_2$ belong to the substrate $\mathcal{S}$ lattice too.
As a consequence, the average corrugation $U(\mathbf{r}_\textrm{c})$ for this class of ``moir\'e-matched clusters'' is independent of the number $N'$ of moir\'e patterns in the cluster and coincides with that of the infinite layer.
Thus, the only surviving components in Eq.~\eqref{eq:Wdefanal} are the CLV with $\delta\mathbf{\Omega}=\mathbf{0}$ that contribute to the corrugation at infinite size: the weight $W$ of any non-CLV vanishes exactly for these ``moir\'e-matched clusters'', as indicated by black arrows in Fig.~\ref{fig:size_match}b.
% NICK: in case of length problems this sentence can go:
The vanishing of $W$ is equivalent to the suppression of the artefact Bragg peaks arising when a non-primitive (e.g., conventional) unit cell is adopted, a well-known feature of the structure factor in crystallography~\cite{Ashcroft1976}, as elaborated in Methods Section~\ref{sec:structure_factor}.
This effect was previously noted in realistic crystalline interfaces \cite{Koren2016,wang2019a}: these ``moir\'e-matched clusters'' exhibit no size effects, as long as they remain perfectly rigid.

For all other parallelepiped clusters of sizes in between these ``moir\'e-matched clusters'', this cancellation does not occur, leading to $W(\delta \mathbf{\Omega},N)\neq 0$ for non CLVs:
the weight function $W$ oscillates as a function of size.
$|W|$ reaches a sequence of maxima, whose peak height decays $\propto N^{-1}$ (blue curve in Fig.~\ref{fig:size_match}b).
Hence the incomplete moir\'e tiles at the edges result in a size-dependent corrugation, as illustrated in Fig.~\ref{fig:size_match}c,d,e.

\subsection{Development of directional structural lubricity} \label{DevStruLub}

Let us now focus on type-B systems, where $\Omega$ is a 1D lattice.
The novelty is that a more restricted class of $\delta \mathbf{\Omega}$'s vanishes, and, as a result, the vast majority of the $\mathbf G$ vectors in the summation of Eq.~\eqref{eqUrN5} lead to size-dependent Fourier contributions.

\begin{figure*}[tb]
\centering
\includegraphics[width=1.0\textwidth]{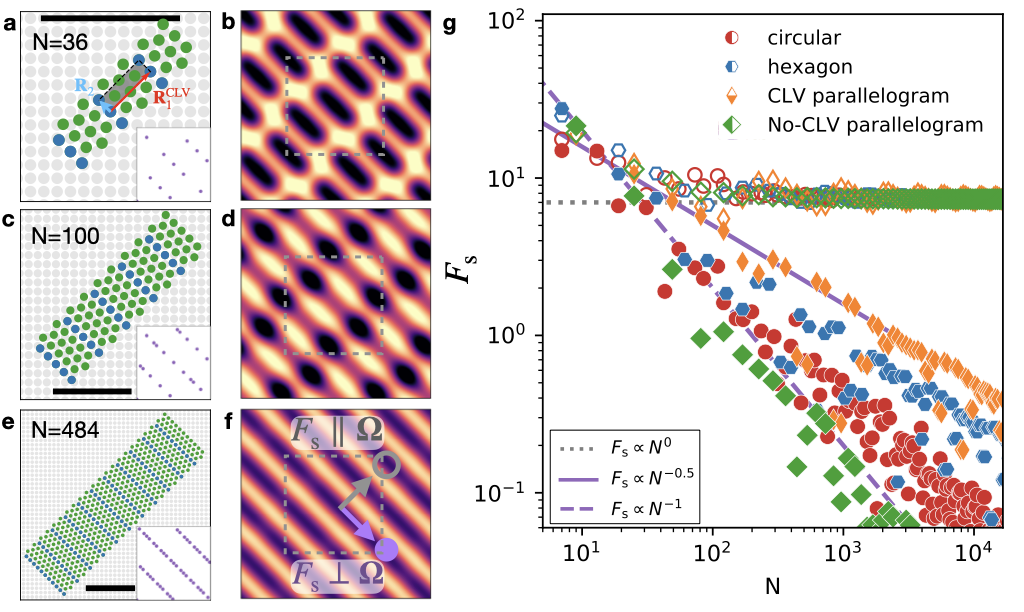}
\caption{
    The development of directional structural lubricity with increasing contact size.
    (a),(c),(e) Adsorbate clusters of different sizes with the same contact incommensurability (type B) as of Fig.~\ref{fig:coverages}(e), namely a triangular-lattice adsorbate with spacing $a=\sqrt{3/2}$ and orientation $\theta_\mathrm{o}=15^\circ$ sliding across a square-lattice substrate with unit lattice spacing.
    These special-shaped clusters here are constructed by repeating a $K=4$-particle unit (shaded area in panel a) based on a real-space CLV $\mathbf{R}^\mathrm{CLV}_1$ (red arrow in panel a) and a primitive vector $\mathbf{R}_2$ (light-blue arrow in panel a).
    The corners of the repeated cell are highlighted in blue.
    The black bar in each panel spans 10 substrate lattice spacings.
    The insets report the corresponding coverage $\mathcal{O}$ for that given size.
    (b), (d), (f) The interlocking potential energy of the clusters in (a), (c), (e) respectively as a function of $\mathbf{r}_\text{c} \in [-1,1]\times[-1,1]$.
    The color code is the same as in Fig.~\ref{fig:coverages}.
    The gray dashed line highlights the Wigner-Seitz cell.
    (g) The cluster-size dependence of static friction for the directionally superlubric interface.
    Filled (empty) symbols refer to the unpinning force parallel (perpendicular) to the low-energy corridors, or perpendicular/parallel to the CLV $\mathbf{\Omega}$, respectively.
    These directions are sketched in panel f.
    The investigated shapes of the clusters (circles, hexagons, parallelograms)
    are reflected in the data-point shapes and colors (red, blue, and orange/green).
    Parallelograms are generated by replicating two kinds of supercells: either based on a real-space CLV (orange) or not (green);
    the adopted supercell lattice vectors are respectively $4\mathbf{R}_1+2\mathbf{R}_2$ and $2\mathbf{R}_2$ (orange), or $2\mathbf{R}_1$ and $2\mathbf{R}_1+2\mathbf{R}_2$ (green), where $\mathbf{R}_1$ $\mathbf{R}_2$ are identified in Fig.~\ref{fig:coverages}e.
    The dotted, solid, and dashed lines report the $F_\mathrm{s} \propto N^0$, $F_\mathrm{s} \propto N^{-1/2}$, and $F_\mathrm{s} \propto N^{-1}$ scalings as guides to the eye.
}
\label{fig:scaling_dirSlub}
\end{figure*}

In the (not guaranteed) circumstance that, beside reciprocal CLVs, we also identify a real-space CLV $\mathbf{R}^\text{CLV}_1$, we can  adopt it as one of the primitive vectors for a supercell.
The second supercell primitive vector $\mathbf{R}_2$ can be chosen freely among the lattice vectors linearly independent of $\mathbf{R}^\text{CLV}_1$, e.g. $\mathbf{R}_2$ in Fig.~\ref{fig:coverages}e.
Compared to the discrete-$\mathcal O$ condition, here $\mathbf{R}_2$ is certainly not a CLV.
As in the previous section, such a supercell contains $K$ vectors $\mathbf{R}_k$, yielding a cluster of size $N=N'K$.
See Fig.~\ref{fig:scaling_dirSlub}a,c,e for examples of clusters built with this protocol.

%\as{For such clusters constructed}, ????
For this class of clusters,
the first factor in Eq.~\eqref{eq:Wdefanal} has $\delta\mathbf{\Omega}\cdot\mathbf{R}^{\mathrm{CLV}}_1=0$,
and thus $\exp(i \delta \mathbf{\Omega} \cdot \mathbf{R}^\mathrm{CLV}_{1}) = 1$, because $\mathbf{R}^\mathrm{CLV}_{1}$ belongs to both the adsorbate and substrate lattice, i.e.\ $\mathbf{R}^\mathrm{CLV}_{1} \in \mathcal{R} \cap \mathcal{S} $.
As a result,
%so that
the second factor leads to a scaling $N^{-1/2}$ for any $\delta\mathbf{\Omega}\neq 0$:
the weight is constant for the CLV $\mathbf{\Omega}$ while the envelope of non-CLV weights decays as $N^{-1/2}$.

The corrugation of this system can be expressed as the sum of the size-independent term, namely the $\mathbf{\Omega}$-sum of Eq.~\eqref{Uofr_linear} (resulting in the corrugation of Fig.~\ref{fig:coverages}h), plus the remaining contributions, which decay as a function of size:
\begin{align}\nonumber
U(\mathbf{r}_\textrm{c}) & = \sum_{n} \tilde{V}_\mathcal{S}(n\mathbf{\Omega}) \exp(i n\mathbf{\Omega} \cdot \mathbf{r}_\textrm{c})
 \\\label{eq:Uofr_linear_correction}
& +  \sum_{\mathcal{G} \textbackslash \Omega} \tilde{V}_\mathcal{S}(\mathbf{G}) \exp(i \mathbf{G} \cdot \mathbf{r}_\textrm{c}) W(\delta\mathbf{\Omega}, N)
\,,
\\\nonumber
W(\delta\mathbf{\Omega}, N) & \propto   K / \sqrt{N}
\,.
\end{align}
An example of size evolution of the energy landscape resulting from Eq.~\eqref{eq:Uofr_linear_correction} is reported in
Fig.~\ref{fig:scaling_dirSlub}b,d,f.
For large size, $U(\mathbf{r}_\textrm{c})$ approaches the straight and flat energy corridors of Fig.~\ref{fig:coverages}h.

The Fourier amplitude of each term in Eq.~\eqref{eq:Uofr_linear_correction} is the product of the weight $W(\delta\mathbf{\Omega}({\mathbf G}), N)$ times the substrate Fourier component $\tilde{V}_\mathcal{S}(\mathbf{G})$ evaluated at the same vector $\mathbf{G}$.
Assuming, as is usually the case, that the surface corrugation potential $V_\mathcal{S}(\mathbf{r})$ is a smooth function, then its Fourier amplitudes
$|\tilde{V}_\mathcal{S}(\mathbf{G})|$
tend to decay with $|\mathbf{G}|$.
As a consequence, long $\mathbf{G}$ vectors even
if
leading to nearly perfect matching (i.e.\ small $\delta\Omega(\mathbf{G})$),
usually yield quite small, practically negligible contributions to the energy landscape.
As a result, the size-dependent term in Eq.~\eqref{eq:Uofr_linear_correction} is usually dominated by a few small-$|\mathbf{G}|$ Fourier components.

This size-dependent energy corrugation has important implication for friction.
In an overdamped context where inertia is negligible, the minimum per-particle force $F_\mathrm{s}$ needed to sustain the motion of the adsorbate crystal in a given direction $\hat{\mathbf u}$, namely the static friction force $F_\mathrm{s}$ in that direction,
%is given by
%$\max_{{\mathbf r}_\mathrm{c}} |
%\hat{\mathbf u} \cdot
%\nabla U({\mathbf r}_\mathrm{c}) |$.
can be
estimated by
% NICK: why not "is evaluated as "? As long as we are allowing any motion perpendicular to u, this is the correct procedure, isn't it?
% AS: there's no motion here, these are all static maps, that's why we thought "estimate" is a more fitting term
% NICK: OK
$\max_{\mathbf{r}_\mathrm{c} \in \mathcal{L}} |
\hat{\mathbf u} \cdot
\nabla U({\mathbf r}_\mathrm{c}) |$,
where $\mathcal{L}$ is the straight line
%trajectory  NICK: suggests some kind of dynamics...
connecting two
%consequent ???
successive
energy minima in the  $\hat{\mathbf u}$ direction.
In the present type-B condition, if $\hat{\mathbf u}$ is aligned along the energy corridors, then only the $\delta\Omega(\mathbf{G})\neq\mathbf 0$ components contribute, leading to $F_\mathrm{s} \propto N^{-1/2}$, whereas if $\hat{\mathbf u}$ has a nonzero component perpendicular to the corridors, then $F_\mathrm{s}$ contains a leading component $\propto N^0$ from the ${\mathbf G} \in \Omega$ Fourier components.

We have verified numerically that the same power laws-scaling of $F_\mathrm{s}$ holds not just for special-shaped parallelogram clusters, but also for different shapes, as reported in Fig.~\ref{fig:scaling_dirSlub}g: similar decays of $F_\mathrm{s}$ parallel to the energy corridor are found for each shape.
However, it is apparent that, for a given size, different cluster shapes can change the value of this parallel friction component by an order of magnitude.
In contrast, the size-independent perpendicular friction component is nearly independent of the cluster shape, too.

These direction-dependent scaling laws %corroborate
justify
the name
{\em directional structural lubricity} for the type-B interface condition:
along the energy corridor the total friction $N\,F_\mathrm{s}$ scales sublinearly with the cluster size, as in standard structural lubricity, while parallel to these corridors the total friction $N\,F_\mathrm{s}$ grows linearly with size, like for a ordinary structurally-matched interface.

\subsection{Close-matching vectors}\label{sec:approxCLV}
We come now to extend the exact classification of Sect.~\ref{sec:infinite_size} to interfaces which --strictly speaking-- belong to type C, but come with a set of
relatively short % note the specification! If we allow G to become arbitrarily big, then it's trivial to find CMVs!  We want G to be rather short, so that the corresponding Fourier component is not negligible...
$\mathbf G$ vectors characterized by a very small mismatch $|\delta\mathbf{\Omega}(\mathbf{G})|$, see Eq.~\eqref{deltaOmega}: close-matching vectors (CMVs).
This approximate classification holds for finite, and not-too-large clusters.

When the two crystals are incommensurate (type-C), the ``standard'' properties of structural lubricity should apply.
However we argue here that in the presence of CMVs the categories and size scaling introduced above survive, up to a maximum cluster size related inversely to $|\delta\mathbf{\Omega}(\mathbf{G})|$.
In this section we focus on directional locking and directional structural lubricity, showing that, in specific size ranges, they are to be expected for type-C interfaces with CMVs.
This analysis is especially relevant for heterocontacts, where perfect matches (whether of type A or B) are unlikely.

We recall that, for each substrate Fourier component identified by $\mathbf{G}$, the $N$-dependence of both factors $f(x)$ in Eq.~\eqref{eq:Wdefanal} implies a critical size below which $W(\delta \mathbf{\Omega}(\mathbf{G}), N) \simeq 1$ because both $f(x)$ factors in $W$ are of order $N^{1/2}$.
For a special-shape cluster (see Section~\ref{specialshapes:sec}) based on the vectors $\mathbf{R}_1$ and $\mathbf{R}_2$, the critical size associated with a substrate vector $\mathbf{G}$ is
\begin{equation}\label{Nc}
N_\mathrm{c} ( \mathbf{G} ) =
\left[\min\left(
\frac{2\pi}{\delta \mathbf{\Omega}( \mathbf{G})\cdot \mathbf{R}_{1} },
\frac{2\pi}{\delta \mathbf{\Omega}( \mathbf{G})\cdot \mathbf{R}_{\mathrm{2}}} \right)\right]^2
,
\end{equation}
where the argument of the square is meant to be rounded to the next integer.
If the cluster size is $N \lesssim N_\mathrm{c}( \mathbf{G})$, then
\begin{equation}
W(\delta \mathbf{\Omega}(\mathbf{G}), N) \simeq 1
\,.
\end{equation}

Let us assume that there exists a single independent CMV $\mathbf{G}' \simeq \mathbf{Q}'$ such that, {\em at its critical size $N^+ =N_\mathrm{c}(\mathbf{G}')$}, $\mathbf{G}'$ gives the dominant contribution to the corrugation energy $U(\mathbf{r}_\mathrm{c})$ of the contact in Eq.~\eqref{eqUrN5}.
The $\mathbf{G}'$ Fourier component becomes the dominating one only beyond some minimum size $N^{-}$, defined as the largest $N < N^+$ such that $\exists \mathbf{G} \in \mathcal{G}\setminus\{\pm \mathbf{G}'\} $ which satisfies $|\tilde{V}_{\mathcal{S}}(\mathbf{G}) W(\delta \mathbf{\Omega}(\mathbf{G}), N)| \ge | \tilde{V}_{\mathcal{S}}(\mathbf{G}') W(\delta\mathbf{ \Omega}(\mathbf{G}'), N)| $.

If the contact conditions are such that $N^-$ is significantly smaller than $N^+$ then this contact exhibits approximate directional superlubricity for all sizes in the range $N^- < N < N^+$.
In this size range, the direction perpendicular to $\mathbf{G}'$ exhibits a very small corrugation associated to minor Fourier components,  negligible compared to the $\tilde{V}_{\mathcal{S}}(\mathbf{G}') W(\delta \mathbf{\Omega}(\mathbf{G}'), N)$ term, which is responsible for a sizeable corrugation in the direction parallel to $\mathbf{G}'$.
As the size $N$ exceeds $N^+$, also this sizeable corrugation perpendicular to the superlubric ``corridor'' begins to fade away due to the decay of $W(\delta \mathbf{\Omega}(\mathbf{G'}), N)$, until ``standard'' direction-independent structural lubricity of an extended type-C contact is recovered.

\begin{figure}[tb]
{
  \centering
  \includegraphics[width=1.0\columnwidth]{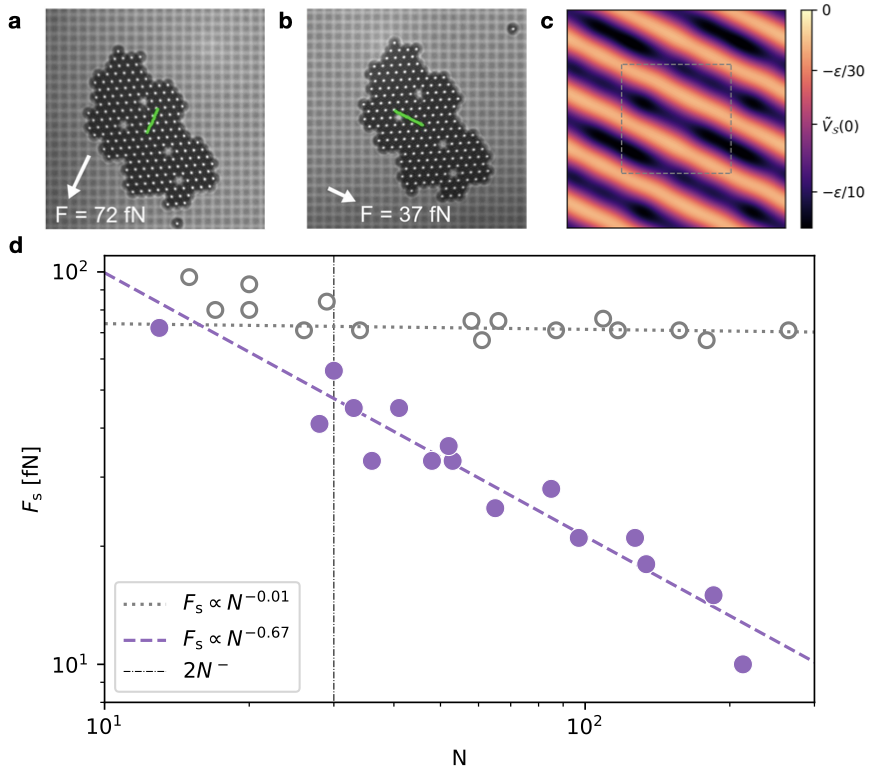}
\caption{
    Experimental realisation of directional structural lubricity. %\xc{XIN: Maybe it's better to switch the order of figure 4 and figure 5. EMA: This is the order in which they appear in the main text, so I think it is required so.}
    (a,b) A triangularly packed colloidal cluster of size $N=185$ with spacing $a=4.45~\mu\textrm{m}$ sliding across a square-lattice surface with spacing $b=5.0~\mu\textrm{m}$.
    %perpendicular and parallel to the energy corridor.
    The green lines in both panels indicate the cluster's center of mass trajectory (see also movie 1) over a period of 85 seconds under the indicated applied forces
    perpendicular (a) and parallel (b) to the energy corridor.
    If the same force strength of 37~fN was applied in the direction of panel (a), the cluster would not move at all:
    a strong friction anisotropy is revealed.
    (c) The calculated potential energy landscape of the cluster in panels (a,b) at orientation $\theta_\mathrm{o}=3.4^\circ$, revealing corridors along the $-25.6^\circ$ direction.
    (d) The experimentally measured static friction force for many different-sized experimental clusters perpendicular (empty gray circles) and parallel (filled purple circles) to the energy corridor.
    Lines are power-law fits $F_\textrm{s} = F_0 N^{\gamma}$  to the corresponding data for clusters of $N>2 N^-\simeq 30$ particles (dot-dash black line).
    The fitted parameters are $F_0=467.4$~fN, $\gamma=-0.67\pm0.07$ and $F_0=76.3$~fN, $\gamma=-0.01\pm 0.02$ for the parallel and perpendicular directions, respectively.
    A detailed description of the experiment is provided in Methods Section \ref{sec:experiment}
    % AS: @Xin please confirm: the clusters rotated at theta_o=+3.4 yield corridors at -25.6 while cluster at theta_o=-3.4 yield corridors at +25.6, right? This is what I see from simulations/theory and what I'd say happens in your experiments from the panel a,b
}
\label{fig:NFc-exp}
}
\end{figure}

We report an experimental test of these predictions, executed letting a triangularly-packed colloidal cluster slide over a surface patterned with a square lattice, as shown in Fig.~\ref{fig:NFc-exp}a,b.
This setup consists of the fully tunable microscale system mimicking an atomistic interface reported previously in Refs.~\onlinecite{Cao2019a, Cao2021, Cao2022}; see Methods~\ref{sec:experiment} for details.
The geometry of the system generates a CMV $\mathbf{G'} \approx \mathbf{Q}_1 - 2\mathbf{Q}_2$ which leads to a nearly-type-B contact
%at small enough
across a broad range of sizes, with clear energy corridors shown in Fig.~\ref{fig:NFc-exp}c.
We can estimate the critical sizes of the system to be $N^- \simeq 15$ and $N^+ \simeq 1620$, adopting a Gaussian model for the
%experimental
corrugation profile of each substrate well, as in Eq.~\eqref{vperiodic}, with parameters taken from Ref.~\onlinecite{Cao2022}.
Figure~\ref{fig:NFc-exp}d reports the measured static-friction forces in the direction of the energy corridors (filled purple circles in Fig.~\ref{fig:NFc-exp}d), and perpendicular to it (empty gray circles in Fig.~\ref{fig:NFc-exp}d), as a function of the cluster size $N$.
The results are in good agreement with the predicted scalings.
Indeed the static friction perpendicular to the energy corridor is approximately constant from $2 N^-$ (dash-dotted line in Fig.~\ref{fig:NFc-exp}d) up to the largest experimental size, with a fitted power-law exponent $\gamma = 0.01 \pm 0.02$ (dotted gray purple line in Fig.~\ref{fig:NFc-exp}d).
On the contrary, the static friction along the energy corridors, %with
exhibits
a power law of exponent $\gamma$ in between $-1$ and $-1/2$ (dashed purple line in Fig.~\ref{fig:NFc-exp}d), in remarkable agreement with the theory given the random shapes of experimental clusters.
%, compatible with the numerical simulations
%{\bf NICK: What numerical simulations??? We haven't described any, just an experiment. Moreover, with the sentence written this way, it seems that the filled symbols are the numerical simulations, no good.  I would remove any mention to numerical simulations, and just write something like "as expected" or even nothing.
%AS: ok it's a bit ambiguous. It was supposed to mean "like numerical results in Fig 4.". But indeed we can save some words and not say anything. It is pretty obvious by looking at fig 4 and fig 5 that the two agree.}
%(filled symbols in Fig.~\ref{fig:scaling_dirSlub}).

\section{Stability against rotation}
\begin{figure}[htb]
  \centering
\includegraphics[width=1.0\columnwidth]{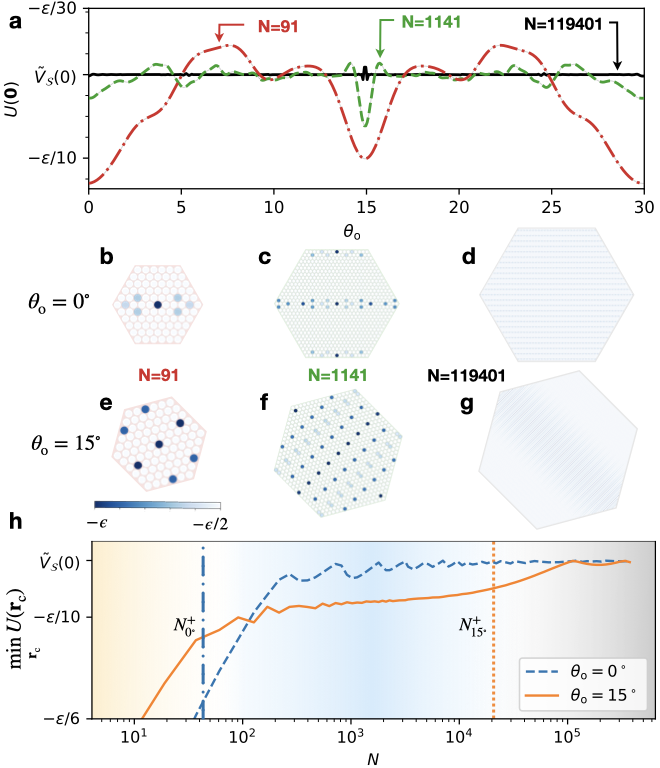}
\caption{ Stability of structural directional lubricity against rotation at finite sizes.
    (a) The interlocking potential Eq.~\eqref{UofrDef} evaluated at fixed $\mathbf{r}_\mathrm{c}=0$ for varying misalignment angle $\theta_\mathrm{o}$ % XIN: IS $\theta_\mathrm{o}$ DEFINED SOMEWHERE? I DID NOT SEE IN ANY FIGURES AT LEAST. EMA: now defined in the text
    for three clusters with size $N=91$, $N=1141$, and $N=119401$, respectively.
    %the center of mass is fixed at $\mathbf{r}_\mathrm{c}=\mathbf{0}$. XIN: I moved this information into the formular $U(\mathbf{r}_\mathrm{c}=0,\theta_\mathrm{o})$
    The contact consists of a triangular-lattice adsorbate spaced by $a = (1+\delta)\sqrt{3/2}$, with $\delta = 0.1\%$ %(i.e. very close to that of Fig.~\ref{fig:coverages}e,f,g,h \xc{XIN: THIS SENTENCE ALREADY APPEARED AT THE MAIN TEXT, DON'T DUPLICATE IT IN FIGURE CAPTION.})
    on a unit square-lattice substrate.
    %The red dot-dashed, green dashed, and black solid lines refer to hexagonal clusters of size $N=91$, $N=1141$, and $N=119401$, respectively. XIN: It is clear enough from the figure which curve corresponds to which size.
    %Note that the \xc{range of $y$-axis, i.e. the} energy scale matches the colorbar in \xc{previous figures. EMA: I don't find it really relevant. Remove?}
    %Figs.~\ref{fig:coverages}d,h,m, \ref{fig:size_match}d--f, \ref{fig:scaling_dirSlub}b,d,f.
    (b--g) Particle-resolved interlocking energy (blue-to-white scale) for the three clusters at two different $\theta_\mathrm{o}$, at fixed $\mathbf{r}_\mathrm{c}=0$.
    %Local potential-energy contributions (blue-to-white scale) for the three clusters oriented at (b--d) $\theta_\mathrm{o}=0^\circ$ or (e--g) $\theta_\mathrm{o}=15^\circ$.
    %{\bf NICK: the labels theta=0 and theta=15 in figure are small to the point of being unreadable. Subscripts missing, too.
    %AS: corrected}
%    White marks particle at energies $V_\mathcal{S}\ge-\epsilon/2$ and dark blue at energy $V_\mathcal{S}=-\epsilon$, as reported in the colorbar below panel e.
% NICK: I added a () so that the sentence above can go.
% NICK: the sentence below (really needed??) is misplaced, this is not where we are discussing panel a!
% AS: I wanted to enphatise that you can directly compare with the other figures, but it's stricly necessary for sure.
%    Note that the energy scale in panel a is the same as the colorbar in Figs.~\ref{fig:coverages}d,h,m, \ref{fig:size_match}d--f,     \ref{fig:scaling_dirSlub}b,d,f.
    (h) The minimum value of $U(\mathbf{r}_\mathrm{c})$ (minimized by allowing all possible center-mass translations $\mathbf{r}_\mathrm{c}$) for fixed alignments $\theta_\mathrm{o}=0^\circ$ (blue dashed) and $\theta_\mathrm{o}=15^\circ$ (orange solid), as a function of size $N$.
    The blue dash-dotted line marks $N_{0^\circ}^+=40$ and the orange dotted line marks $N_{15^\circ}^+ = 17684$.
    %{\bf NICK: the fact that you need to specify left and right indicates that you need to use different dashing (e.g. dotted and dot-dashed) for these 2 vertical lines. Also, please note that $r_c$ at the vertical-scale label should be bold not italic (2 spots).
    %AS: corrected}
}
\label{fig:dirSlub_rotation-hex}
\end{figure}

So far, the relative orientation of the two crystals, has been implicitly fixed by the
%relative alignment
angle
%{\bf orientation} % NICK: I have no idea who wrote this in bold.  Anyway, we certainly do not want bold text in a paper!  As for the word, it seems an unneeded repetition: I add "angle" and make it more specific:
$\theta_\mathrm{o}$
%of
between
the first primitive vectors of the two lattices, {$\mathbf{R}_1$ and $\mathbf{S}_1$}.
Clearly, upon rotation the same system may realize a type-A contact (2D array of CLVs), or a type-B contact (1D CLV array), or a type-C contact (fully incommensurate, no CLVs).
In practice it is unlikely that one can artificially keep the contact at an arbitrarily fixed mutual angle: in most concrete setups the contact will eventually relax to an energetically stable condition.
It is therefore essential to examine the angular energetics of such contacts and their stability upon rotation.

For example, it is well known that structural lubricity in homocontacts arises from the misalignment of the two  lattices.
However, this misalignment comes with an energy cost, that makes the superlubric contact unstable~\cite{dienwiebel2004,Filippov2008,deWijn2010}.
In homocontacts therefore energetics acts to stabilize the type-A geometry.

We argue that for heterocontacts the same stabilization occurs.
Depending on the geometric details, this stabilization may lead to either a type-A or a type-B contact.
Under such conditions, the important outcome is that {\em directional locking and directional structural lubricity are energetically stable}.
If the  crystals mutually rotate from a fully incommensurate kind-C geometry where the interlocking energy is given by Eq.~\eqref{Uofr_dense}, to such an angle that CLVs in the reciprocal space arise, additional terms appear in the
interlocking potential
of Eqs.~\eqref{Uofr}, leading to potentials of the form \eqref{Uofr_discrete} or \eqref{Uofr_linear}.
% AS: unclear wording, re-phrased below
%Consequently, for those orientations, the corrugated Fourier components provide energy to be gained
%for the center-mass positions $\mathbf{r}_\textrm{c}$ which minimize the corrugation energy.
%from the relaxation of the center-mass position $\mathbf{r}_\textrm{c}$ to stable equilibrium point.
Consequently, for these orientations, these Fourier components in the interlocking potential provide an energy lowering at the equilibrium position, not available in type-C configurations.
This means that when a contact allows for type-A or type-B conditions, the corresponding orientations are indeed the most stable ones.

It is straightforward to check this energetics numerically,
not only for the infinite contacts of Eq.~\eqref{Uofr}, but also for finite-$N$ cluster of Eq.~\eqref{eqUrN5}.
As an illustration, we select a system quite close to the type-B one of Fig.~\ref{fig:coverages}e,f,g,h
(triangular-lattice adsorbate over a square substrate), but with a small mismatch $\delta = 0.1\%$ introduced in the adsorbate lattice spacing $a = (1+\delta)\sqrt{3/2}$.
Due to the small $\delta$, at relative orientation $\theta_\mathrm{o}=15^{\circ}$ the CLV $\mathbf{\Omega}$ of Fig.~\ref{fig:coverages}f turns into a CMV $\mathbf{G}'=2\pi(2,2)$, with a corresponding critical size $N^+_{15^{\circ}} \simeq 1.8 \times 10^4$.
To study the relative stability of the consequent nearly-type-B contact, this orientation has to be compared with all others.
Figure~\ref{fig:dirSlub_rotation-hex}a reports the potential energy $U(\mathbf{r}_\mathrm{c} = \mathbf{0})$ of Eq.~\eqref{UofrDef} as a function of the misalignment angle $\theta_\mathrm{o}$.
The energy profiles of Fig.~\ref{fig:dirSlub_rotation-hex}a indicate an evident local energy minimum at $\theta_\mathrm{o}=15^{\circ}$ for $N=91$, which becomes the global minimum for $N=1141 < N^+_{15^{\circ}}$.

For sizes $N \lesssim 100$ the global minimum is found for a different orientation, $\theta_\mathrm{o}=0^{\circ}$.
This second minimum corresponds to a different (shorter, but worse matched) CMV $\mathbf{G}''=2\pi(0,1)$, with critical size $N^+_{0^{\circ}} \approx 40$.
The moir\'e patterns associated with these two CMVs at $\theta_\mathrm{o}=0^\circ$ and $\theta_\mathrm{o}=15^\circ$ are shown for different sizes in Fig.~\ref{fig:dirSlub_rotation-hex}b,c,d and Fig.~\ref{fig:dirSlub_rotation-hex}e,f,g, respectively.
To illustrate the relative stability between these two orientation, Fig.~\ref{fig:dirSlub_rotation-hex}h reports the equilibrium interlocking energy as a function of size.
%This condition reveals that at small size $N \lesssim 100$ the alternative stable
This alternative $\theta_\mathrm{o}=0^{\circ}$ orientation is also a nearly-type-B contact.
The direct comparison between the CMV at the two orientations of $\tilde{V}_S(\mathbf{G}')W(\delta\mathbf{\Omega}(\mathbf{G}', N))$ and  $\tilde{V}_S(\mathbf{G}'')W(\delta\mathbf{\Omega}(\mathbf{G}''), N)$ as a function of size gives a crossover point at $N \simeq 100$, which coincides to the crossing point in Fig.~\ref{fig:dirSlub_rotation-hex}h, above which $\theta_\mathrm{o}=15^\circ$ becomes the equilibrium orientation.

At larger sizes, $N \gg N^+_{15^{\circ}}$, no CMVs retains a significantly large Fourier component in Eq.~\eqref{Uofr} and, as expected for a type-C contact, the energy profile as a function of orientation becomes nearly flat, as in the  $N = 119401$ example of Fig.~\ref{fig:dirSlub_rotation-hex}a.

These observations indicate that directional locking and directional structural lubricity are not just a hypothetical eventuality.
On the contrary, they prove that with the condition that at some orientation $\theta_\mathrm{o}$ the contact geometry generates a CLV, or even just a CMV, sufficiently short to be associated to a sizeable corrugation Fourier component, then precisely this Fourier component is responsible for the energy stabilization of this orientation, which %is then the one that
the contact will reach spontaneously if %it is
allowed to realign.

Indeed, in the colloidal experimental realizations reported here in Sec.~\ref{sec:approxCLV} and in Refs.~\onlinecite{Cao2019a, Cao2021} where the clusters are free to reorient, directional locking phenomena and reorientation emerge spontaneously as the result of self-alignment and not of external manipulation.

\section{Directional structural lubricity in real-life contacts}\label{real-life:sec}

\begin{figure*}[tb]
  \centering
\includegraphics[width=1.0\textwidth]{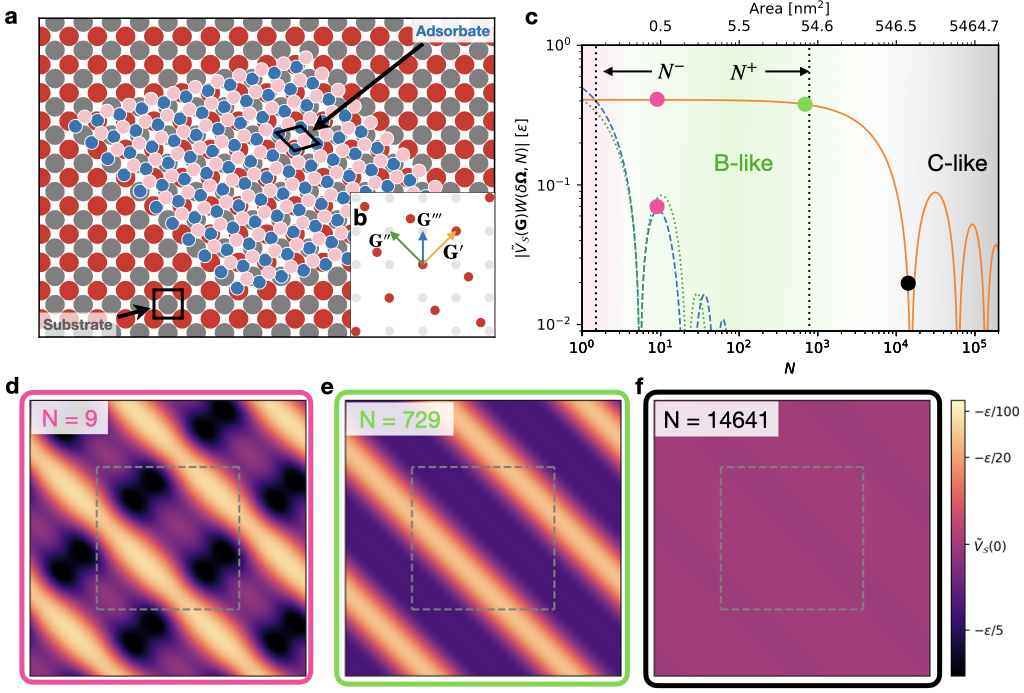}
\caption{
    A realistic interface with approximate directional structural lubricity.
    (a) A monolayer hBN flake (blue dots=B atoms; pink dots=N atoms) deposited on a VO(001) surface (gray dots=V atoms; red dots=O atoms) substrate.
    The angular misalignment is $\theta_\mathrm{o}=15^\circ$.
    (b) Reciprocal lattices of the adsorbate ($\mathcal{Q}$, red) and substrate ($\mathcal{G}$, gray),
    %with marked a CMV $\mathbf{G}'= (1,1)2\pi/b$ (orange arrow),
    with marked a CMV $\mathbf{G}'$ (orange arrow),
    a perpendicular reciprocal vector %$\mathbf{G}_\mathrm{P}= (-1,1)2\pi/b$ (green arrow),
    $\mathbf{G}''$%= (-1,1)2\pi/b$
    (green arrow),
    and the vector %$\mathbf{G}_{2}= (0,1)2\pi/b$
    $\mathbf{G}'''$%= (0,1)2\pi/b$
    (blue arrow) yielding a sizable Fourier component at small flake size.
    (c) The size dependence of the Fourier components associated to
    $\mathbf{G}'$ (orange solid line),
    $\mathbf{G}''$ (green dotted),
    %$\mathbf{G}_\mathrm{P}$ (green dotted),
    and $\mathbf{G}'''$ (blue dashed).
    %and $\mathbf{G}_2$ (blue dashed).
    The vertical lines mark the $\mathbf{G}'$ critical sizes $N^-\simeq 1$ and $N^+ = 779$ defined in Sect.~\ref{sec:approxCLV}.
    (d,e,f) Maps of the interlocking potential $U(\mathbf{r}_\mathrm{c})$ in the $[-b,b]\times[-b,b]$ square for the three selected flake sizes marked as matching-colored circles in panel (c);
    the energy scale is indicated at the right.
    The dashed gray square delimits the Wigner-Seitz cell of the substrate lattice $\mathcal{S}$.
}
\label{fig:trasl_en-real_mater}
\end{figure*}

By taking advantage of CMVs, we can identify interfaces of real materials that should exhibit approximate directional structural lubricity across significant size ranges.

Consider the system depicted in Fig.~\ref{fig:trasl_en-real_mater}a: for the adsorbate we take a flake of hexagonal boron nitride (hBN), a layered material with hexagonal symmetry and spacing $a=\SI{0.2512}{nm}$~\cite{Jain2013a}; for the substrate we adopt the (001) surface of VO~\cite{dellaNegra2000VO}.
This surface has
square symmetry and spacing $b=\SI{0.310}{nm}$~\cite{Jain2013a}.

When the hBN flake is rotated at $\theta_\mathrm{o}= 15^\circ$ as
in Fig.~\ref{fig:trasl_en-real_mater}a, the interface exhibits a CMV $\mathbf{G}'=2\pi/b(1,1)$ (orange arrow in Fig.~\ref{fig:trasl_en-real_mater}b),
that fosters an energetically stable configuration.
The two next most significant Fourier components are associated to  %$\mathbf{G}_\mathrm{P}$
$\mathbf{G}''= 2\pi/b\,(-1,1)$
 (perpendicular to $\mathbf{G}'$) %(green arrow)
and  %$\mathbf{G}_\mathrm{2}$ (blue arrow).
$\mathbf{G}'''= 2\pi/b \, (0,1)$.

To illustrate the effect at hand, here we adopt a radically simplified energy landscape,  obtained by summing the attraction of each N and B atom in the adsorbate with every substrate atom (regardless of it begin V or O), represented by a negative Gaussian function $V(\mathbf{r})$, with a straightforward extension of Eq.~\eqref{vperiodic}.
The resulting interlocking potential depends on two parameters only: the width $\sigma = 0.1\,b$ of the Gaussian function, and its peak attraction $\epsilon$, which we keep undefined, and adopt as the energy scale of this example, thus expressing all energies in Fig.~\ref{fig:trasl_en-real_mater} in terms of $\epsilon$.
Of course, a realistic force field would imply quantitatively different Fourier components $V_\mathcal{S}(\mathbf{G})$, but we do not expect the results to
%be modified
change
radically, because the size-dependent weights $W(\delta\mathbf{\Omega}(\mathbf{G}'), N)$ would be identical.
Note that $N$ indicates the number of lattice cells, consistently with the rest of the paper.
In the hBN flake the total number of atoms is $2N$.

Figure~\ref{fig:trasl_en-real_mater}c reports the size dependence of the Fourier amplitude $|V_\mathcal{S}(\mathbf{G}') \, W(\delta\mathbf{\Omega}(\mathbf{G}'), N)|$ (solid orange line), plus the analogous quantity for $\mathbf{G}''$ (dotted green) and $\mathbf{G}'''$ (dahsed blue).
The $\mathbf{G}'$ component dominates across the size range from $N^- \simeq 1$ up to a critical size $N^+ = 779$.
As a result, at small size, multiple substrate $\mathbf G$ vectors  contribute significantly to the interlocking potential, resulting in a relatively irregular landscape dominated by pronounced energy corridors modulated by a secondary weaker corrugation, as exemplified in Fig.~\ref{fig:trasl_en-real_mater}d for $N=9$.
This secondary corrugation associated mainly to $\mathbf{G}''$  and $\mathbf{G}'''$ decays rapidly with increasing size, see Fig.~\ref{fig:trasl_en-real_mater}c.
Across the size range $N^-\ll N<N^+$ spanning over two orders of magnitude in area, the
$\pm\mathbf{G}'$ %\as{AS: isn't it a bit confusing this $\pm$ here?}
Fourier components remain effectively the dominant contribution to the interlocking potential, with the result that this interface exhibits approximate directional structural lubricity as exemplified in Fig.~\ref{fig:trasl_en-real_mater}e for $N=729$.
For sizes larger than $N^+$, the energy landscape flattens out and the infinite-size limit of ordinary (kind-C) structural lubricity is approached, as exemplified in Fig.~\ref{fig:trasl_en-real_mater}f for $N=14641$.

The hBN/VO(001) interface is just an example where we predict directional structural lubricity to arise.
Another interface where it could be observed is WSe$_2$ on CuF(001), as discussed in
%Methods Section~\ref{sec:real_mater2},
SI Section 1,
and others can be discovered by going through existing materials databases \cite{Jain2013a,Mounet2018,Bergerhoff1987,Kirklin2015OQMD}.

\begin{figure*}
    \centering
    \includegraphics[width=\textwidth]{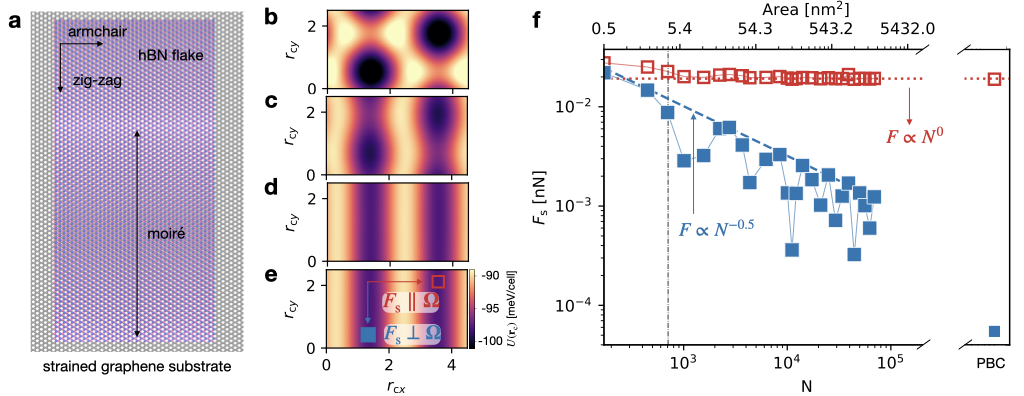}
    \caption{
    Directional structural lubricity via strain engineering.
    (a) An aligned hBN flake on a graphene substrate strained by $\varepsilon_\mathrm{armchair} \simeq 1.8\%$ in the armchair=$\hat x$ direction.
    The aspect $x:y$ ratio of the hBN flake is approximately $1:2$.
    The wavelength of the linear moir\'e pattern is highlighted with a black arrow.
    (b-e) Corrugation potential $U(\mathbf{r}_\mathrm{c})$ for hBN flakes of size
    (b) 6.5 nm $\times$ 12.7 nm ($N=170$),
    (c) 13 nm $\times$ 25.5 nm ($N=2760$),
    (d) 39 nm $\times$ 76.5 nm ($N=11120$),
    (e) PBC
    in the region $[0, 0.45~\text{nm}\times [0, 0.25~\text{nm}]$.
%    Note that as
    Like in Fig.~\ref{fig:trasl_en-real_mater}, $N$ is the number of lattice cells, so that the flake consists of $2N$ atoms in total.
    %The colorbar in panel e reports the scale of $U(\mathbf{r}_\mathrm{c})$ in meV per hBN cell.
    %{\bf NICK: I would replace the previous sentence with a units label meV/cell near the lowest colorbar only. I invite you to adopt the SAME meV range from -99.9 to -89, and the same 5 meV interval for all 4 plots (i.e. ticks at -95 and -90). Whatever goes outside this range, to hell, let it saturate!
    %AS: done}
    (f) Static friction in the armchair lattice-matched direction (red empty symbols) and zig-zag (blue filled symbols), respectively parallel and perpendicular to the CLV $\mathbf{\Omega}$.
    The dashed (dotted) line indicates the scaling in the direction parallel (perpendicular) to the energy corridors, as indicated in panel e.
    The vertical dash-dotted black line marks the critical size $N_\mathrm{c}$ of the largest non-matching component $\mathbf{G}'$.
    }
    \label{fig:hBN_Gstrained}
\end{figure*}
In addition, the geometric conditions for directional structural lubricity can be achieved by means of strain engineering, a method that has  been recently used to tailor frictional
%properties~\cite{Androulidakis2020strain,zhang2019tuning,zhu2019c}.
properties~\cite{Androulidakis2020strain,zhang2019tuning,Wang2019strain}.
For example, the well-studied structurally lubric hBN/graphite contact can be modified by a uni-axial strain applied to the graphite substrate in the
%zig-zag
armchair
direction, as shown in Fig.\ref{fig:hBN_Gstrained}a.
This deformation, at
%$\epsilon_\text{zig-zag}=1.8\%$,
%$\varepsilon_\text{armchair} = 1.81799\%$,
$\varepsilon_\mathrm{armchair} \simeq 1.8\%$, i.e. within experimental feasibility~\cite{cao2020elastic},
would generate the geometrical conditions for a type-B directional structural lubricity along the
%armchair
zig-zag
direction as discussed in Methods Section~\ref{sec:MD}.
On such a strained graphene surface, we evaluate the static-friction unpinning threshold for aligned hBN flakes of different sizes, based on a realistic force field~\cite{Brenner2002,Leven2016,kinaci2012thermal}.

Figure~\ref{fig:hBN_Gstrained}f reports our prediction for this static-friction force in two orthogonal directions: perpendicular to the valleys of the interlocking potential, i.e.\ in the strained
%zig-zag
armchair
direction, and parallel to them, i.e.\ in the unstrained
%armchair
zig-zag
direction, as depicted in Fig. \ref{fig:hBN_Gstrained}e.
Like in the colloid experiment of Fig.~\ref{fig:NFc-exp}, the
%simulation % AS: since these are not really MD simulations...
numerical
results agree with the expectation for a type-B contact:
in the directions perpendicular to the valleys $F_\text{s} \propto N^0$, while parallel to the valleys $F_\text{s} \propto N^{-1/2}$.

To probe the effect of elasticity, we allow for relaxation of the atomic  positions, both of the hBN cluster and of the graphite substrate, while controlling the respective center-of-mass positions.
We adopt periodic boundary conditions (PBC), which unavoidably introduce a secondary, much weaker strain
%for graphite in the perpendicular direction
on the zig-zag edges.
%$\varepsilon_\mathrm{zig-zag}=0.00294\%$, applied to the hBN layer. % AS: already in Methods.
%{\bf NICK: Ehm, this is all fine, but probably the detail of the strain value and that we applied it to hBN can be moved to the methods part. What is probably more important to note here is:
This small strain leads to a commensurate (type-A) configuration, that fits in a periodic cell.
This forced commensuration is responsible for the tiny, but nonzero friction in the zig-zag direction reported in the rightmost point in Fig.~\ref{fig:hBN_Gstrained}f.
%(2) Differently from finite clusters, overall size or shear adjustments (that elasticity allows and that might lead to other period matchings, different from the imposed one) are suppressed.

As detailed in the Methods Section~\ref{sec:MD}, at each center-mass position $\mathbf{r}_\text{c}$, we perform %quasi-static NICK: I can't see what this "quasi-static" specification conveys.  The COM is static all right, not quasi.  The atoms move to where they want in order to minimize the energy: they are not static.  So I would avoid this misleading "quasi static" name.
a systematic evaluation of the total interlocking energy and lateral force
after a full atomic relaxation.
The friction force
%components
parallel to the valleys
after relaxation
%are approximately one order of magnitude
is nearly a factor three
larger than when they are evaluated for rigid layers as reported in Table~\ref{tab:MD_F_anisotropy}.
%in Methods Section~\ref{sec:MD}.
Despite this reduction, these results
%do confirm the large
indicate that elastic deformations preserve the substantial
friction anisotropy.
%{\bf NICK: the following sentence seems useless at this point.  These numbers are in the table: if the reader is interested (s)he goes there to read them...}
%The ratio of the static friction components in the two perpendicular directions $F_\mathrm{s}^\mathrm{armchair}/F_\mathrm{s}^\mathrm{zig-zag}$ is 347 and 140 for rigid and flexible contacts, respectively.

%\as{AS: Mention that even with strain not exactly 1.81799 you should still see this behaviour, as pointed out in hBN/VO(001) example.} NICK: I did:
The adopted value of strain leads to an exact type-B contact.  Small deviations from that value would lead to a type-C contact that, for not-too-large hBN clusters, would lead to an approximate type-B behavior, as discussed in Sect.~\ref{sec:approxCLV}.
% AS: great thanks!

\section{Discussion}
In this work we formulate a precise classification of the matching/mismatching conditions of 2D crystalline contacts.
In addition to the well-known lattice-matching condition and to the structurally lubric fully mismatched condition, we discover an intriguing intermediate condition, characterized by vanishing static friction in just one special direction.
We extend this rigorous theory from the rather abstract domain of infinite lattices to the approximate but experimentally more relevant situation of finite-size contacts, where we provide estimations of the range of validity of the resulting frictional regimes.

This theory, summarized in Eqs.~\eqref{Uofr_discrete} and
\eqref{Uofr_linear}, shows how both directional locking and structural lubricity can occur in non-trivial preferential directions, i.e.\ directions that generally do not coincide with those of highest symmetry for either the substrate or the adsorbate.
This is possible only when the dominant contributions to the interlocking potential originates from higher Fourier components $\mathbf{\Omega}$s of the corrugation potential.
Non-trivial directions are consequently absent for purely sinusoidal potentials, as was noted in Ref.~\onlinecite{deWijn2012}.
Furthermore, the existence of directional structural lubricity is possible only when adsorbate and substrate have either low symmetries or different symmetries (see Methods~\ref{sec:rotation_arg}).
This excludes the commonly studied cases of homocontacts, and even heterocontacts of triangular-on-triangular and square-on-square lattices, which perhaps explains why directional structural lubricity has gone unnoticed so far.
Regarding this novel frictional regime, we provide concrete examples of its realization, including an experiment based on colloidal particles driven across a patterned surface.

Most importantly, we investigate the angular energetics of the problem, and show that the same contact orientations that support directional locking and directional structural lubricity are the most energetically stable.
%
% AS: We say better in Section VI and I think here is not relevant to the point, and actually breaks the flow.
%Notably, the angular stability of the directional structurally lubric contact originates from precisely the same mechanism stabilising the lattice-matched, statically-pinned configurations of type-A, i.e. tendency to gain interaction energy from matching Fourier components.
Hence the well-known spontaneous decay of superlubricity in homocontacts \cite{Filippov2008} does not apply to directional structural lubricity.

Precisely this intrinsic stability could suggest applications of directional structural lubricity in contexts where a dramatic friction anisotropy is required.
The natural field of application involves nanoparticles/nanocontacts, whose small size can accommodate the not-quite-perfect CMVs that one is likely to encounter in real life.
By tuning temperature and taking advantage of the different thermal expansion of the contacting materials, or by applying
strain to one of the crystals, as discussed in Sec.~\ref{real-life:sec},
precise CLVs can be achieved:
this condition allows one to realize perfect directional structural lubricity even for a macroscopically large contact, as long as elastic effects are negligible. % AS: AV felt this note was needed to make this claim of "macroscopically" less misleading and introducing the next paragraph

%An important point to note is that t
The present investigation does not include
%the effects of
elasticity.
For rigid materials, the effect of
%intra-bond relaxations????
the elastic response
should
%be reasonably negligible
remain small and even negligible
%for several systems,
as long as the contact size remains below a critical value.
Such critical sizes range from
%$N=9\times 10^4$
$N\simeq 10^5$ % NICK order of magnitude seems sufficient here, especially since we do not give details, and below you approximate even more radically!...
for the colloidal clusters of Fig.~\ref{fig:NFc-exp}, to $N\simeq 10^{12}$
%{\bf NICK: as usual here N is the number of cells, please insert in the right order of magnitude, probably something like 12 or 13...
%AS: 1mm^2/0.5nm^2=1e12nm^2/0.5nm^2~1e12}
corresponding
to %lengths of millimeters
linear dimensions in the millimeter region
for hBN/strained graphene (see Methods~\ref{sec:MD}) and MoS$_2$/graphene heterostructures \cite{Liao2022,Cao2022}.
%, up to centimeters for double-walled carbon nanotube~\cite{Ma2015}.  NICK: I fail to see how the DWCNT can fit in our 2D contact picture. Perhaps we better leave them out. AS: fair point, was there from the rebuttal letter of PRX.
For softer materials, or larger sizes,
elasticity will
acquire an increasingly important role,
%detrimental ???
usually leading to higher friction
\cite{Sharp2016,varini2015islands}.
For type-C contacts, the Aubry-transition paradigm
\cite{aubry1983FKanal,Braun1998,Brazda2018a}
leads us to predict that structural lubricity
%gets lost
gives way to a high-friction regime
as soon as the strength of the adsorbate-substrate interaction starts to prevail over the adsorbate and substrate rigidity.
%Interestingly, it is not clear to us whether softness would always hinder the phenomena presents in quasi-type-B and quasi-type-A contacts \textit{a priori} where CMVs are present. Elasticity may on the contrary lead to the emergence of local solitonic excitations with corresponding wave-vectors $\delta \Omega$s, as happens for the case of Novaco-McTague misalignments \cite{novaco1977}, which at finite size could enhance the anisotropy of friction which leads to directional structural lubricity.
% NICK: I don't believe the sentence above. Please go back to read https://journals.aps.org/prl/abstract/10.1103/PhysRevLett.114.108302 : the novaco physics ENHANCES friction...
% I try to rewrite:
For soft type-B and near-type-B contacts, we expect a regular 1D Aubry-type physics, although other effects related to Novaco-McTague distortions \cite{novaco1977} might play a role too \cite{mandelli2015frictionNovaco, mandelli2017finite, Brazda2018a}.
These questions however go beyond the scope of this paper, and as such their investigation is left to future work.

\section*{Acknowledgement}
A.S.\ thanks D. Kramer and A.\ de Wijn for the useful discussions.
X.C.\ acknowledges funding from Alexander von Humboldt Foundation.
E.T.\ acknowledges support by ERC ULTRADISS Contract No. 834402.
N.M., A.V., and A.S.\  acknowledge support by the Italian Ministry of University and Research through PRIN UTFROM N. 20178PZCB5.

\makeatletter
\setcounter{equation}{0}

\section*{Author contributions statement}
E.P., A.S., and N.M.\ derived the mathematical formulation.
X.C.\ carried out the experiments.
A.S.\ and E.P.\ wrote the computer code for the rigid simulations.
A.S.\ and J.W.\ performed the numerical simulations.
All authors contributed to the theoretical understanding, discussed the results and wrote the paper.

\section*{Competing Interests}
The authors declare no competing interests

\section*{Data and Code Availability Statements}
Data used in this work is available from the corresponding author on reasonable request.

\section*{Materials and Methods}
\section{Properties of the
discrete-coverage dual and real-space lattices} \label{sec:corollary_rational}
In the discrete-coverage condition, summarized in row A of Table~\ref{tab:conditions}, the following two statements hold:
(i) for all $\mathbf{Q}_j \in \mathcal{Q}$, their components $\mathbf{Q}_{j\mu}/(2\pi) \in \mathbb{Q}$ ($\mu=x,y$), i.e. they are rational numbers;
(ii) likewise, for all $\mathbf{R}_j \in \mathcal{R}$, their components $\mathbf{R}_{j\mu} \in \mathbb{Q}$, rational numbers too.

The demonstration of (i) goes as follows:
we can express any lattice vector in $\mathcal{Q}$ as a linear combination of the primitive lattice vector $\mathbf{Q}_a$, $\mathbf{Q}_b$, with integer coefficients.
In particular, for the lattice vectors $\mathbf{Q}_1$, $\mathbf{Q}_2$
whose components $\mathbf{Q}_{i\alpha} /(2\pi)$ are integers (line A of Table~\ref{tab:conditions}),
\begin{align}
    \mathbf{Q}_1 &= n_1 \mathbf{Q}_a + n_2 \mathbf{Q}_b \label{eq:coroll11}\\
    \mathbf{Q}_2 &= m_1 \mathbf{Q}_a + m_2 \mathbf{Q}_b, \label{eq:coroll12}
\end{align}
with $n_1, n_2, m_1, m_2 \in \mathbb{Z}$.
The relations in Eq.~\eqref{eq:coroll11}, \eqref{eq:coroll12} can be inverted to express the primitive vectors in term of $\mathbf{Q}_1, \mathbf{Q}_2$:
\begin{align}
    \mathbf{Q}_a &= \frac{m_2 \mathbf{Q}_1 - n_2 \mathbf{Q}_2}{n_1 m_2-n_2 m_1} \label{eq:coroll21}\\
    \mathbf{Q}_b &= \frac{-m_1 \mathbf{Q}_1 + n_1 \mathbf{Q}_2}{n_1 m_2-n_2 m_1}. \label{eq:coroll22}
\end{align}
Similar relations hold for these vectors divided by $2\pi$:
\begin{align}
    \frac{\mathbf{Q}_a}{2\pi} &= \frac{m_2 \frac{\mathbf{Q}_1}{2\pi} - n_2 \frac{\mathbf{Q}_2}{2\pi}}{n_1 m_2-n_2 m_1} \label{eq:coroll31}\\
    \frac{\mathbf{Q}_b}{2\pi} &= \frac{-m_1 \frac{\mathbf{Q}_1}{2\pi} + n_1 \frac{\mathbf{Q}_2}{2\pi}}{n_1 m_2-n_2 m_1}. \label{eq:coroll32}
\end{align}
Since at the right hand side of Eqs.~\eqref{eq:coroll31},
\eqref{eq:coroll32} all vectors have integer components, the primitive vectors $\mathbf{Q}_a/(2\pi), \mathbf{Q}_b/(2\pi)$ have rational components.
As an arbitrary lattice vector $\mathbf{Q}_j\in \mathcal Q$ can be written as an integer-coefficient combination of these primitive vectors, $\mathbf{Q}_j = l_1 \mathbf{Q}_a + l_2 \mathbf{Q}_b $ (with $l_1,l_2 \in \mathbb Z$), we conclude that all lattice points in $\mathcal{Q}$ divided by $2\pi$ have rational components.

The demonstration of the real-lattice statement (ii) goes as follows: given the primitive vectors $\mathbf{Q}_a$, $\mathbf{Q}_b$ of $\mathcal{Q}$, then the primitive vectors of $\mathcal R$ can be obtained through the following formulas~\cite{Ashcroft1976}:
\begin{align}
\mathbf{R}_a &= 2\pi \frac{R_{90} \cdot \mathbf{Q}_b}{\mathbf{Q}_a \cdot R_{90} \cdot \mathbf{Q}_b}
= \frac{R_{90} \cdot \frac{\mathbf{Q}_b}{2\pi }}{\frac{\mathbf{Q}_a}{2\pi } \cdot R_{90} \cdot \frac{\mathbf{Q}_b}{2\pi }}
\\
\mathbf{R}_b &= 2\pi \frac{R_{90} \cdot \mathbf{Q}_a}{\mathbf{Q}_b \cdot R_{90} \cdot \mathbf{Q}_a}
% NICK: The denominator is the area of the reciprocal-space primitive cell.  I don't see why we should write it differently for this case. See next passage
= \frac{R_{90} \cdot \frac{\mathbf{Q}_a}{2\pi }}{\frac{\mathbf{Q}_a}{2\pi } \cdot R_{90} \cdot \frac{\mathbf{Q}_b}{2\pi }}
\,,
\end{align}
where $R_{90}$ represents the $2\times2$ $90^\circ$ rotation matrix.
The rightmost  expressions involve only rational quantities, which proves that $\mathbf{R}_a$ and $\mathbf{R}_b$  have rational components.
As a consequence, all vectors in $\mathcal R$ have rational components.

Finally, by multiplying $\mathbf{R}_a$ and $\mathbf{R}_b$ by the least common denominator of their respective (rational) components, one readily obtains two independent vectors in $\mathcal R$ characterized by all integer components.

\section{Necessary condition for line coverage}
\label{sec:rotation_arg}

To allow for the possibility of line coverage (type B), the two lattices must not share a rotational symmetry of order $n>2$: in practice they cannot be both square or triangular lattices.
To prove this, consider two such lattices sharing a rotational symmetry by an angle $\alpha$ of order $n>2$.
First of all, if they do not have any nonzero CLV
%vector????]
in reciprocal space then they
%would be
are
in the dense coverage case (type C). %, by construction.
Let us then assume that they do have a nonzero CLV
%vector ?????
$\mathbf{\Omega}^*$.
As a consequence of the common symmetry, also the rotated vector $R_{\alpha}(\mathbf{\Omega}^*)$ is a CLV  (where $R_{\alpha}$ represents a rotation by $\alpha$).
Moreover, if the symmetry order is larger than $2$, then $\mathbf{\Omega}^*$ and $R_{\alpha}(\mathbf{\Omega}^*)$ are both CLVs and independent, which leads by definition to the case of discrete coverage, type A.
We conclude that to have line coverage (type B) either the two lattices must have different symmetries, or share the same low-order symmetry.

\section{Weight function structure factor}\label{sec:structure_factor}
When, as is usually the case, $\mathbf{R}^\text{CLV}_1$ and $\mathbf{R}^\text{CLV}_2$ are not primitive vectors of $\mathcal{R}$, they define a supercell, namely the unit cell of the moir\'e pattern formed by $\mathcal R$ and $\mathcal S$.
This supercell contains $K$ vectors $\mathbf{R}^k$, such that any lattice-translation vector $\mathbf{R}_j$ defining the cluster can now be identified as $\mathbf{R}_j = j_1 \mathbf{R}^\text{CLV}_1 + j_2 \mathbf{R}^\text{CLV}_2 + \mathbf{R}^k$, with $j_1,j_2 = -(\sqrt{N'}-1)/2,\dots,(\sqrt{N'}+1)$, $k = 1, \dots, K$, and we assume that $\sqrt{N'}$ is an odd integer such that $N=N'\,K$.

We then express the weight function as \begin{align}\nonumber
W(\delta \mathbf{\Omega}, N) =&
\frac 1N \! \sum_{j_1, j_2, k}
\!\!\!\exp(i\delta \mathbf{\Omega}\cdot (j_1 \mathbf{R}^\text{CLV}_1+ j_2 \mathbf{R}^\text{CLV}_1 + \mathbf{R}^k))
\\\nonumber
 =& \frac{1}{N}
\frac{\sin(\sqrt{N'} \delta \mathbf{\Omega}\cdot \mathbf{R}^\text{CLV}_1 / 2) }{ \sin( \delta \mathbf{\Omega}\cdot \mathbf{R}^\text{CLV}_1 / 2)}
\times
\\\label{eq:WwithS}
& \quad \frac{ \sin(\sqrt{N'} \delta \mathbf{\Omega}\cdot \mathbf{R}^\text{CLV}_2 / 2) }{ \sin( \delta \mathbf{\Omega}\cdot \mathbf{R}^\text{CLV}_2 / 2)} \, S(\delta \mathbf{\Omega})
\,.
\end{align}
Here we have introduced the supercell "structure factor"
\begin{align}\label{eq:Sdef}
S(\delta \mathbf{\Omega})
=\sum_{k=1}^K
\exp(i\delta \mathbf{\Omega}\cdot  \mathbf{R}^k)
\,.
\end{align}
The main observation here is that since $\mathbf{R}_i^{\mathrm{CLV}}$ are CLV belonging to both $\mathcal{R}$ and $\mathcal{S}$, this structure factor vanishes exactly for all nonzero $\delta \mathbf{\Omega}$'s.
By definition $\delta\mathbf{\Omega}\cdot \mathbf{R}_i^{\mathrm{CLV}}=\mathbf{G}\cdot\mathbf{R}_i^{\mathrm{CLV}}-\mathbf{\bar{Q}}\cdot\mathbf{R}_i^{\mathrm{CLV}}=2\pi q$ with $q\in \mathbb{Z}$.
Hence, the sinc terms in eq. \eqref{eq:WwithS} are equal to $1$ and the weights $W$ associated with any substrate vector $\mathbf{G}$ are size-independent.
But only CLV contributions can be size independent and survive for infinite monolayer.
Thus the non-CLV contribution must vanish, hence $S(\mathbf{\delta\Omega})=0$.

\section{Experimental details}\label{sec:experiment}
The data shown in Section \ref{fig:NFc-exp} are obtained using the experimental apparatus described in~\cite{Cao2019a, Cao2021}, where monolayers of triangularly packed colloidal crystalline clusters of up to hundreds of particles in size are firstly created and then driven across a periodic (square lattice) potential under an applied force.
To form the colloidal clusters, we inject a colloidal suspension into a sample cell of $20\times30\times0.3~\mathrm{mm}^3$ in size, where the $0.3~\mathrm{mm}$ is the sample thickness and the $20\times30~\mathrm{mm}^2$ is the sample area.
The colloidal suspension contains a dilute amount ($\sim10^7/\mathrm{mL}$) of colloidal particles (Dynabeads M450 with a diameter of $a=4.45~\mu\mathrm{m}$) in a water-based solution that contains a small amount of polyacrylamide (0.02\% by weight) and sodium dodecyl sulfate (50\% critical micelle concentration). The polyacrylamide induces a bridging flocculation effect which causes the colloidal particles to attract strongly when they get close to each other. Due to gravity, the colloidal particles (buoyant weight $mg=286~\mathrm{fN}$) sediment on the bottom surface of the sample cell.
Under Brownian motion, the colloidal particles on the sample surface meet one another and aggregate to form random-shaped small clusters up to several tens of particles in size.
To facilitate the formation of larger clusters, we tilt the sample so that the small clusters can move quickly on the sample surface and grow larger via accretion.
This process also clears the sample surface so that during future sliding the clusters will hardly bump into each other.

The sample surface contains portions of periodically corrugated regions created by photolithography.
To create the periodic structures, a thin layer (thickness 100 nm) of photoresist (SU8 2000) was firstly coated to the sample surface and then exposed by UV light (wavelength 365~nm) under a photomask that contains the predefined periodic patterns.
The surface is then washed in SU8 developers, after which the unexposed part of the thin film dissolve away, forming a periodically corrugated structure on the surface. This creates a periodic potential for the colloidal particles with a lattice spacing $b=5~\mu\mathrm{m}$ and potential barrier $25.8~k_\mathrm{B}T_\mathrm{room}$.

Due to the orientational locking effect,
%as observed in~\cite{Cao2021}
during sliding the clusters will become orientationally locked to an angle $\theta_\mathrm{o}=-3.4^{\circ}$ relative to the periodic surface's lattice direction. At this angle a CMV appears at $\mathbf{G'} \approx \mathbf{Q}_1 - 2\mathbf{Q}_2$ which leads to a nearly-type-B contact.
This creates a low energy corridor along the $25.6^{\circ}$ direction on the interlocking potential energy landscape of the clusters.

%We then
To apply a driving force $F$, we
tilt the sample in such a way that the in-plane component of the gravitational force
%, i.e. the driving force $F$
is either parallel to the low energy corridor or perpendicular to it.
To
%obtain
determine
the static friction force component $F_\mathrm{s}$ reported in Fig.~\ref{fig:NFc-exp}, we firstly increase $F$ %towards
in the
80--100~fN region to ensure that the cluster can move along or perpendicular to the low-energy corridor.
We then gradually lower $F$ until the cluster is no longer moving: this is the measured $F_\mathrm{s}$.

\section{Simulating directional structural lubricity of hBN on strained graphite}\label{sec:MD}
% AS: NOTE it's graphiete, since we have the zsprings
%{\bf The MD simulation
%NICK: what we report here, is no MD simulation at all... In the rigid calculations we are just using lammps to evaluate the U function.  In the relaxation calculations we are doing minimizations, which can hardly be called "MD"...  I reformulate:
%AS: I agree, well done.}
% JIN: Yes, I agree with the suggestion. Not "dynamics" at all.
We simulate a realistic model for the hBN/strained graphene interaction,
consisting of a finite-size rectangular hBN slider in contact with a
graphene substrate
%\as{AS: we do have vertical springs mimicking the rest of the bulk, so perhaps graphite is more appropriate?}
to which PBC are applied along the
%zig-zag ($x$)
armchair ($x$)
and
%armchair ($y$)
zig-zag ($y$)
directions, see Fig.~\ref{fig:hBN_Gstrained}a.
%To achieve the
%single??? THERE ARE INFINITELY MANY CLV's, THEY FORM A 1D LATTICE....
To obtain a contact with
%the single
a
CLV $\mathbf{\Omega}=(4.625, 0)~\mathrm{nm}^{-1}$
%needed for ??? Type B does not especially require this specific CLV!
compatible with
directional structural lubricity
along the zigzag direction
, we strain the graphene substrate (with original bond length $1.42039$~\r{A})
along the
%zig-zag
armchair
direction by
$\varepsilon_\mathrm{armchair}=(a_\mathrm{hBN}/b_\mathrm{graphene}-1) = 1.81799\%$
to match the hBN lattice
%(bond length $1.44214$~\r{A})
(bond length $1.44621$~\r{A}), while along the zig-zag
direction the graphene substrate is shrunk by
$\varepsilon_\mathrm{zig-zag}=-0.34542\%$
to account for the Poisson ratio $\nu=0.19$ \cite{Jain2013a}.
%{\bf NICK: as x is the zig-zag direction, the CC bonds are not aligned along x. Therefore a strain in the x direction by 1.8\% is determined by an elongation of the bonds forming the zig zags by a bit more than 1.8\%. Or by a change in the angles, as can occur due to a shrinkage in the y direction.  As we do have this, the shortening of these zig-zag bonds can be more or less than 1.8\%.
%Anyway, to provide complete info about our geometry we need to specify the lengths of the CC bonds in the zig-zag direction, of those directed along y and one angle.
% I guess they are given by the equilibrium condition of the REBO potential...
%In general we should give these bond-length specifications after those of strain, approximately here:}
The strained graphene has armchair-directed bonds of length 1.44621~\r{A},
zig-zag bonds of length 1.42323~\r{A}, and the angle between them amounts to $120.5357^\circ$.
The negative stress required for this elongation is estimated to be 8 GPa, assuming a Young modulus of graphene of 1TPa.
This is within the experimentally achievable strain of graphene~\cite{cao2020elastic}.
For all simulations of finite-size hBN clusters, both hBN and graphene are kept rigid.
The hBN-graphene interaction is described by the accurate registry-dependent inter-layer potential (ILP) \cite{Leven2014}.
In this system the
%substrate
reciprocal
vector yielding the next most significant Fourier component of the interlocking potential is $\mathbf{G}'=(2.312, 4.076)~\mathrm{nm}^{-1}$.

We also simulate the infinite-size layer, by applying PBC to both hBN and graphene in a common supercell of size
%$10.02~\text{nm}\times 19.96~\text{nm}$,
$5.64~\text{nm}\times 11.5~\text{nm}$ ($N=1196$),
that imposes a minimal residual strain of 0.00294\% to hBN in the
%armchair
zig-zag
direction.

\begin{table}[tb]
    \centering
    \begin{tabular}{c|c|c}
    Force &  Rigid & Flexible \\
    component &  [nN] & [nN] \\
\hline
    %$F_{\mathrm{s}x}$ &  $2.76 \times 10^{-04}$ & $1.20 \times 10^{-03}$
    $F_{\mathrm{s}}^\mathrm{armchair}$ &  $1.90 \times 10^{-2}$ & $2.20 \times 10^{-2}$ \\
    %$F_{\mathrm{s}y}$ &  $2.50 \times 10^{-06}$ & $2.45  \times 10^{-05}$
    $F_{\mathrm{s}}^\mathrm{zig-zag}$ &  $ 5.46\times 10^{-5}$ & $ 1.58 \times 10^{-4}$\\
    \hline
    $F_{\mathrm{s}}^\mathrm{armchair}/F_{\mathrm{s}}^\mathrm{zig-zag}$& 347 & 140
    \end{tabular}
    \caption{
    Static friction components for a hBN infinite monolayer (simulated with PBC) dragged along strained graphite.
    The first column reports the static friction force per
    %atom
    cell
    %{\bf NICK: please correct to per hBN cell, coherently with the figure and the rest of the paper. AS: Done
    %NICK: the caption still mentions "per atom": please Andrea check carefully, verify that all is coherent!
    %AS: we changed the code, so the numbers are right, just missed this in the caption, sorry.}
    for a rigid layer (rightmost points in Fig.~\ref{fig:hBN_Gstrained}f) and the second column  the same quantity evaluated for the flexible case.    \label{tab:MD_F_anisotropy}
    }
\end{table}
The ratio of friction between the pinned armchair direction and the directionally structurally lubric direction is rather large, as expected.
Beside simulating rigid layers, in order to probe the effect of elastic deformation, in this PBC model we perform additional flexible simulations, with $z$-direction springs tethered to each slider and substrate atom to mimic the elasticity of the bulk materials \cite{guo2012mechanics}.
The average external load is kept to zero during the sliding.
%{\bf NICK: the other attachment point of these springs fixes the average load applied to this interface.  What load are we considering?  We should tell... AS: it's zero load, written now.}
When modeling the elastic infinite-size contact, both the slider (hBN) and the substrate (graphene) are fully flexible and periodically repeated.
The intralayer interaction of hBN and graphene are described by shifted-Tersoff \cite{sevik2011characterization,ouyang2019mechanical,mandelli2019princess} and REBO \cite{Brenner2002} force fields, respectively.
A standard (quasi-static) simulation protocol \cite{bonelli2009atomistic, mandelli2017sliding} is adopted to compute the interlocking potential energy $U(\mathbf{r}_\text{c})$
while changing the position $\mathbf{r}_\text{c}$ of the slider relative to the graphene substrate, whose center-of-mass (COM) is kept fixed throughout the simulation.
The COM $\mathbf{r}_\mathrm{c}$ of the slider is scanned on a $0.1$~\r{A} grid over $x$ and $y$.
%\textcolor{red}{or $\delta y=0.1 $\r{A} and ended up scanning a 3 \r{A} $\times$ 4 \r{A} region}.
%{\bf NICK: the U function has of course the same periodicity as the strained graphene.  It is sufficient to explore one primitive cell of that lattice, and convenient to scan it in steps that are an integer fraction of the cell size.  Anyway, even 0.1\AA{} is fine, of course, just a little wasted time... JIN: Yes. Technically just need to scan a primitive cell region and use dx and dy as (ax, ay)/10. Even though it doesn't make big difference to Fs, I am sorry about the mis-chosen parameters...}
%The fix COM is then fixed ???? AS: added
At fixed COM, the structure is  relaxed until the force experienced by each atom decreases below $10^{-5}$~eV/\r{A}.
All calculations of the interface potential energy for the rigid layers, and its relaxation in the elastic case are conducted by means of the open-source LAMMPS code \cite{LAMMPS22}.

After obtaining the energy field $U(\mathbf{r}_\mathrm{c})$, we estimate the static friction $F_\mathrm{s}$ along the expanded
%zig-zag ($\hat{u}_\mathrm{zz}=\hat{x}$)
armchair ($\hat{u}_\mathrm{armchair}=\hat{x}$)
%and weakly shrunk armchair ($\hat{u}_\mathrm{arm}=\hat{y}$)
and weakly shrunk zig-zag ($\hat{u}_\mathrm{zig-zag}=\hat{y}$)
direction by taking the  maximum value of %$F_{\mathrm{zz}}=-\hat{u}_\mathrm{zz}\cdot \nabla U$ and
$F^{\mathrm{armchair}}=-\hat{u}_\mathrm{armchair}\cdot \nabla U$ and
%$F_{\mathrm{arm}}=-\hat{u}_\mathrm{arm} \cdot \nabla U$.
$F^{\mathrm{zig-zag}}=-\hat{u}_\mathrm{zig-zag} \cdot \nabla U$ between two successive minima, as in Section~\ref{DevStruLub} and Fig.~\ref{fig:scaling_dirSlub}g.
Table \ref{tab:MD_F_anisotropy} compares the static friction obtained taking elastic displacement into account with that obtained for the rigid layer.

The overall anisotropy and thus the armchair/zigzag friction ratio, while  still very large, is reduced by elasticity.
Note that the finite size of the supercell of PBC calculations effectively implements a small-wavevector cutoff, that forbids all long-wavelength deformation.
The size at which long-wavelength elastic deformations become important can be estimated by the critical length defined by Sharp {\it et al.}\ \cite{Sharp2016}:
\begin{align}
    \lambda=Gd/\tau,
\end{align}
where $G$ is the in-plane shear modulus of the material, $d$ is lattice constant of the substrate, and $\tau$ is the interface shear strength.
For the case of hBN on graphite, we obtain a critical length
$\lambda=0.662~\mathrm{mm}$, using the following values from the literature:
$d=0.246~\mathrm{nm}$ \cite{Brenner2002},
$\tau=0.12~\mathrm{MPa}$ \cite{Song2018RobustSuperlub},
$G=2G_\mathrm{graphene} G_\mathrm{hBN}/(G_\mathrm{graphene}+G_\mathrm{hBN})$~\cite{johnson1987contact},
with $G_\mathrm{graphene}=372~\mathrm{GPa}$ and $G_\mathrm{hBN}=285~\mathrm{GPa}$ \cite{Jain2013a}.

\end{document}